\documentclass[12pt]{iopart}

\usepackage{graphicx, epsfig} 
\usepackage{bm}
\usepackage[export]{adjustbox}

\usepackage[caption=false]{subfig}
\usepackage[usenames]{xcolor}
\usepackage{cite}

\definecolor{linkcolor}{rgb}{0.0, 0.3, 0.5}
\definecolor{purple}{rgb}{0.7, 0.05, 0.5}
\definecolor{robin_lilac}{rgb}{0.8,0.6,0.9}
\usepackage[unicode,colorlinks=true,citecolor=linkcolor,linkcolor=linkcolor,urlcolor=linkcolor, hyperfootnotes=true,hyperfootnotes=true]{hyperref}

\usepackage[all]{hypcap}
\expandafter\let\csname equation*\endcsname\relax
\expandafter\let\csname endequation*\endcsname\relax
\usepackage{amsmath,amsfonts,amssymb}
\usepackage[normalem]{ulem}
\usepackage{orcidlink}
\usepackage{etoolbox}

\newcounter{count}

\newcommand{\du}{\mathrm{d}}
\newcommand*{\defeq}{\mathrel{\vcenter{\baselineskip0.5ex \lineskiplimit0pt
\hbox{\scriptsize.}\hbox{\scriptsize.}}}
                      =}

\usepackage{array}
\usepackage{cancel}
\usepackage{wasysym}
\newcolumntype{C}[1]{>{\centering\let\newline\\\arraybackslash\hspace{0pt}}m{#1}}

\usepackage[utf8]{inputenc}
\usepackage{footmisc}
\usepackage{enumitem}

\makeatletter
\newcommand{\mainmatter}{%
  \setcounter{footnote}{0}%
  \patchcmd{\@makefntext}{\fnsymbol}{\arabic}{}{}%
  \patchcmd{\@thefnmark}{\fnsymbol}{\arabic}{}{}%
  \def\@makefnmark{\textsuperscript{\arabic{footnote}}}%
}
\makeatother

\begin{document}
{\hfill \scriptsize KCL-PH-TH/2022-59}

\title[Unequal-mass boson-star binaries: Initial data and merger dynamics]{Unequal-mass boson-star binaries: Initial data and merger dynamics}

\author{
Tamara Evstafyeva \orcidlink{0000-0002-2818-701X} $^1$, 
Ulrich Sperhake \orcidlink{0000-0002-3134-7088} $^{1,2,3}$, 
Thomas Helfer \orcidlink{0000-0001-6880-1005} $^{2}$, 
Robin Croft \orcidlink{0000-0002-1236-6566} $^{1,4}$, 
Miren Radia \orcidlink{0000-0001-8861-2025} $^1$, 
Bo-Xuan Ge \orcidlink{0000-0003-0738-3473} $^5$, 
Eugene A. Lim \orcidlink{0000-0002-6227-9540} $^5$
}

\address{$^1$ Department of Applied Mathematics and Theoretical Physics, Centre for Mathematical Sciences, University of Cambridge, Wilberforce Road, Cambridge CB3 0WA, United Kingdom}

\address{$^2$ Department of Physics and Astronomy, Johns Hopkins University, 3400 N Charles Street, Baltimore, Maryland 21218, USA}

\address{$^3$ Theoretical Astrophysics 350-17, California Institute of Technology, 1200 E California Boulevard, Pasadena, CA 91125, USA}

\address{$^4$ Dipartimento di Fisica, “Sapienza” Università di Roma \& Sezione INFN Roma1, Piazzale Aldo Moro 5, 00185, Roma, Italy}

\address{$^5$ The Department of Physics, King's College London, The Strand, London WC2R 2LS, United Kingdom.}

\ead{te307@cam.ac.uk, U.Sperhake@damtp.cam.ac.uk}
\vspace{10pt}

\begin{abstract}
We present a generalization of the curative initial data construction
derived for equal-mass compact binaries in Helfer {\it et al} (2019 Phys. Rev. D 99 044046; 2022 Class. Quantum Grav. 39 074001) to arbitrary mass ratios.
We demonstrate how these improved initial data avoid
substantial spurious artifacts in the collision dynamics of
unequal-mass boson-star binaries in the same way as has previously
been achieved with the simpler method restricted to the equal-mass case.
We employ the improved initial data to explore in detail the impact
of phase offsets in the coalescence of equal- and unequal-mass boson
star binaries.
\end{abstract}

\vspace{2pc}
\noindent{\it Keywords}: Numerical Relativity, Gravitational Waves, Boson Stars

\maketitle
\mainmatter

\section{Introduction}

Perhaps no concept is more central to modern physics than that of the field. Fields are building blocks of our universe and they play a key role in most
paradigms of modern cosmology and theories that extend the Standard Model (SM) of particle physics. In recent years, inflationary \cite{Amin:2014fua,Amin:2011hj,Amin:2010dc} and dark matter (DM) models \cite{Borsanyi:2015cka, Marsh:2015wka} have given an important role to scalar fields, which also naturally arise from string theory \cite{Svrcek:2006yi}. If given mass, scalar fields coupled to gravitational field can theoretically form astrophysical, compact, star-like objects. One of the examples of such stars include \textit{boson stars} (BSs) 
described by a complex, massive scalar field; see \cite{Liebling:2012fv, Schunck:2003kk} for a review. The constituents of a BS are bosonic particles, or bosons (hence, the name), whose mass in the range of $10^{-22} - 10^{-3} eV$ has been considered in cosmological and astrophysical settings \cite{DiGiovanni:2022xcc, Marsh:2015xka}. 
While the existence of fermionic compact objects -- such as neutron stars
or white dwarfs --  is supported by a plethora of observational evidence
(e.g.~\cite{Shklovsky1967,Richer:1997jk,LIGOScientific:2021qlt}),
the search for localized solitons made of bosons is still ongoing. 

Self-gravitating bosonic fields were first studied in the form
of Wheeler's {\it gravitational electromagnetic entities} or {\it geons}
\cite{Wheeler:1955zz}.
The concept of BSs, in the sense of equilibrium solutions to the
Einstein equations, followed about a decade later with
Kaup's pioneering work \cite{Kaup:1968zz}
on self-gravitating configurations of massive complex scalar field, dubbed as \textit{Klein-Gordon geons}. Originally these configurations were devised from fundamental scalar (spin 0) fields \cite{Ruffini:1969qy, Feinblum:1968nwc} and later on extended to vector (spin 1) fields (aka \textit{Proca stars}) \cite{Brito:2015pxa,Minamitsuji:2018kof,Brito:2015yga,Zhang:2021xxa,March-Russell:2022zll,Gorghetto:2022sue,Herdeiro:2021lwl,Minamitsuji:2017pdr,Sanchis-Gual:2017bhw,Duarte:2016lig,SalazarLandea:2016bys,Zilhao:2015tya} or high-spin fields \cite{Jain:2021pnk,Jain:2022nqu}. The nature of the scalar field can be real --
resulting in potentially long lived but not strictly stable
compact objects commonly referred to as
{\it oscillatons} (OSs) \cite{Hawley:2002zn, Helfer:2016ljl, Muia:2019coe, Urena-Lopez:2012udq} -- or complex for BSs; this latter complex
case is the focus of our work. The first calculations of BSs employed free massive scalar fields, resulting in so-called {\it mini boson stars}. Extensive studies over the years, however,
have uncovered a rich variety of other BS models, most
notably through more elaborate scalar potential functions:
self-interacting \cite{Colpi:1986ye, Schunck:1999zu}, solitonic \cite{Friedberg:1986tq, Boskovic:2021nfs} or axionic potentials \cite{Urena-Lopez:2012udq, Guerra:2019srj, DiGiovanni:2022xcc}.
The self-interaction terms lead to significantly more compact BSs,
comfortably exceeding the compactness of neutron stars, and also
increase the maximum mass BSs may acquire without forming a black hole (BH)
\cite{Colpi:1986ye,Lee:1986ts}.
Further BS models include charged stars \cite{Jetzer:1989av, Pugliese:2013gsa}, BSs comprised of multi-fields \cite{Alcubierre:2018ahf, Alcubierre:2021psa} and rotating models \cite{Kleihaus:2005me, Schunck:1996he}, where the nature of the spin is quantised. The stability of various bosonic configurations has been addressed in Refs.~\cite{Kleihaus:2011sx, Sanchis-Gual:2019ljs, Siemonsen:2020hcg} and numerous numerical relativity (NR) simulations have demonstrated the robustness of the models \cite{Muia:2019coe, Balakrishna:2007mr, Bezares:2022obu, Croft:2022bxq}.

Due to their (potentially) very high compactness, BSs belong to
the category of \textit{exotic compact objects} and are even
regarded as candidates for ultracompact {\it BH mimickers} in the sense
of possessing a light ring \cite{Cardoso:2016rao,Maggio:2021ans}. 
More generally, thanks to their comparatively simple and mathematically regular nature but rich
phenomenology, BSs are ideal proxies to study fundamental properties of 
compact objects using analytic and numerical methods.
In this spirit, BSs are also intriguing probes in our search for evidence
of extra degrees of freedom in theories of gravity extending general relativity (GR).
From an observational viewpoint, BSs have been suggested as alternatives to primordial BHs \cite{Mielke:2000mh} and supermassive BHs in the centres of galaxies \cite{Torres:2000}. Last but not least, BSs may contribute
to the dark-matter sector of the universe and
are an important target for gravitational-wave (GW) observations with the LIGO-Virgo-KAGRA (LVK)
network \cite{Berti:2018cxi, Cardoso:2017cqb, Sennett:2017etc, Bustillo:2020syj, CalderonBustillo:2022cja},
as well as future
detectors like the Einstein Telescope and Laser Interferometer Space Antenna (LISA) \cite{Shaddock:2008zz, Maggiore:2019uih}.

Searches for BS signatures with these GW detectors require
accurate waveform models \cite{Chia:2020psj,Toubiana:2020lzd}
whose construction, in turn,
relies on extensive high-precision numerical simulations
of binary systems involving BSs. The numerical exploration of
orbiting BS binary systems is still a relatively young field, but 
has already demonstrated the potentially rich phenomenology of
the GW signals generated by these systems. To our knowledge, the first
investigation dates back to Palenzuela {\it et al} \cite{Palenzuela:2007dm} who find that
the BSs' phase offset affects the merger phase more strongly than
the inspiral.
The GW signal generated by the merger remnant is furthermore mainly
governed by the fundamental oscillation frequency of the remnant
as it settles down into a non-spinning configuration \cite{Palenzuela:2017kcg}. 
Quite remarkably, the GW signal
from BS binary mergers can be exceptionally long-lived, resulting
in a GW {\it afterglow} that decays at a much slower rate than
the remnant's angular momentum \cite{Croft:2022bxq}. The inspiral
of unequal-mass BS binaries has been studied in Ref.~\cite{Bezares:2022obu} and can result in large kicks of thousands of km/s which, however, is due to the asymmetric ejection of bosonic matter rather than that of GWs.
So-called {\it dark boson-star binaries} with purely
gravitational interaction have been found to generate
GW signatures
distinguishable from other astrophysical objects like black holes, neutron stars and even ``normal'' BSs \cite{Bezares:2018qwa}.
Further simulations of compact binaries involving BSs include the
piercing of bosonic clouds by a BH \cite{Cardoso:2022vpj} and
the inspiral of neutron stars with bosonic dark cores
\cite{Bezares:2019jcb}.

In spite of the tremendous progress made in these numerical explorations,
our understanding of the GW emission across the BS binary parameter
space remains very limited, both in terms of coverage and precision.
One key ingredient indispensable for the systematic construction of GW
waveforms forms the central focus of this paper: the generation of
accurate initial data representing plausible physical configurations with negligible
violations of the Einstein constraint equations. The importance of initial data for binary BS star evolutions in the equal-mass case has been previously addressed by Helfer et al.~\cite{Helfer:2018vtq, Helfer:2021brt}, who demonstrate how inaccurate initial superposition of boson stars can lead to substantial spurious features in the resulting gravitational waveforms; to overcome these issues they further propose a new binary superposition that we dub the \textit{equal-mass fix}. This binary initial data is also applied to the case of equal-mass binary {\it neutron star} initial data in the FUKA code \cite{Papenfort:2021hod}. We see here an example how BS studies
serve as a valuable proxy well beyond the immediate scope of BS physics. A key limitation of the above cure, however, is its restriction to
equal-mass binaries. In this work, we develop a generalized version
of this method that achieves the same benefits for binaries with arbitrary mass ratios and contains
the equal-mass fix as a limiting case in the choice of two free parameters. The proposed methodology can be applied to head-on collisions as well as systems with angular momentum, as for example in Ref.~\cite{Croft:2022bxq}. Here we employ this improved initial data construction in
the simulation of equal- and unequal-mass BS binary head-on collisions
studying systematically the impact of the BSs' phase offset on the
collision dynamics and resulting GW signals.

The outline of this work is as follows. We start by introducing the theoretical framework and the BS model of interest in Section \ref{sec:framework}. In Section \ref{sec:3plus1} we summarise the $3+1$ split of our equations of motion and the code infrastructures used for our simulations. Section \ref{sec:initial-data} opens with a brief review of the improved construction of initial data in the equal-mass case
and proceeds with the generalisation to unequal mass ratios. We explore the parameter space of this initial data construction in Section \ref{sec:parameter_space}. The results from our exploration of the
BSs' phase parameter are presented in Section \ref{sec:results} and we conclude in Section \ref{sec:conclusion}. Throughout this work, we set $\hbar = c = 1$ and use $M$ to denote the total mass of the binary. We label spacetime indices by Greek letters running from 0 to 3 and spatial indices by Latin indices running from 1 to 3.

\vspace{1cm}

\section{Theoretical framework} \label{sec:framework}
Mathematically, boson stars (BSs)
are localized, soliton-like\footnote{A soliton describes a wave packet solution that maintains its shape during propagation.} solutions of the coupled system of the Einstein and general relativistic Klein-Gordon equations for a complex scalar field $\varphi$. The action is given by the Einstein-Hilbert term for four-dimensional gravity and a minimally coupled complex scalar field
\begin{equation} \label{eq:action}
    S = \int \sqrt{-g} \left\{\frac{1}{16 \pi G} R - \frac{1}{2} \left[g^{\mu \nu} \nabla_{\mu} \bar{\varphi} \nabla_{\nu} \varphi + V(\varphi) \right] \right \} d^4x,
\end{equation}
where $V(\varphi)$ is the scalar potential. Varying this action, we recover the Einstein and matter evolution equations
\begin{align}
    G_{\alpha \beta} &= 8 \pi G T_{\alpha \beta}, \\
    \nabla^{\mu} \nabla_{\mu} \varphi &= 
    \frac{\du V}{\du |\varphi|^2},
\end{align}
and the energy-momentum tensor reads
\begin{equation}
    T_{\alpha \beta} = \partial_{(\alpha}\bar{\varphi}\partial_{\beta)}\varphi - \frac{1}{2}g_{\alpha \beta}\left[g^{\mu \nu}\partial_{\mu} \bar{\varphi} \partial_{\nu} \varphi + V(\varphi) \right].
\end{equation}
In this work we will focus on the solitonic potential first proposed in \cite{Lee:1986tr}
\begin{equation} \label{eq:potential}
    V_{\rm{sol}} = \mu^2 |\varphi|^2 \left(1 - 2\frac{|\varphi|^2}{\sigma_0^2} \right)^2,
\end{equation}
where $\mu$ is the mass of the scalar field and $\sigma_0$ quantifies the field's self-interaction. Note that the solitonic potential has multiple roots in $|\varphi|$: $|\varphi| = 0$, which corresponds to the true vacuum state and $|\varphi| = \sigma_0/\sqrt{2}$, which represents a "false" or "degenerate" vacuum state. The potential \eqref{eq:potential} can result in highly compact stars and allows us to span a wider range of mass ratios. Furthermore, solitonic potentials produce some particularly interesting solutions; for example, in the case of $\varphi \sim \sigma_0/\sqrt{2}$ thin-wall configurations have been found, where the scalar field profile acquires a shape almost like a Heaviside function \cite{Collodel:2022jly}. The resulting soliton profile is then split into three different regions: the interior solution where $\varphi \sim \sigma_0/\sqrt{2}$ (i.e. a false vacuum state), a transition region with a sharp drop from $\varphi \sim \sigma_0/\sqrt{2}$ to $\varphi = 0$ and the exterior true vacuum state $\varphi = 0$.

In this work, we focus on time evolutions of head-on BS collisions. In general, the outcome of the collision is a non-spinning boson star or a black hole. However, a scenario where the two BSs “pass through” each other \cite{Choi:2009} is also possible. In our head-on collisions, the resulting remnant is always a BH. We model single BSs as stationary solutions in spherical symmetry, where our ansatz splits the complex solution into amplitude $A(r)$, constant frequency $\omega \in \mathbb{R}$ and phase-offset $\delta \phi \in [0,2\pi)$ 
\begin{equation} \label{eq:ansatz}
    \varphi(r,t) = A(r) e^{i (\omega t + \delta \phi)}.
\end{equation}
With this ansatz we construct single equilibrium BS solutions using a shooting algorithm. For details of this construction see Section 2.3 of Ref. \cite{Helfer:2021brt}.
\section{The 3+1 decomposition and computational infrastructure} \label{sec:3plus1}
The simulations of BS collisions in this work have been performed
with two independent numerical relativity codes,
{\sc grchombo} \cite{Andrade:2021rbd,Radia:2021smk}
and {\sc lean} \cite{Sperhake:2006cy}. Both codes evolve the Einstein-Klein-Gordon equations using conformal variants of the 3+1 formalism of Arnowitt, Deser and Misner (ADM)
\cite{Arnowitt:1962hi} as reformulated by York
\cite{York1979}; cf.~also \cite{Gourgoulhon:2007ue}. 
Here, the spacetime metric is decomposed into the spatial metric $\gamma_{ij}$,
the shift vector $\beta^i$ and the lapse function $\alpha$
in adapted coordinates $x^{\alpha}=(t,x^i)$ according to
\begin{equation}
  \du s^2 = g_{\mu\nu}\du x^{\mu}\du x^{\nu}
  =
  -\alpha^2 \du t^2
  +\gamma_{mn}
  (\du x^m + \beta^m \du t)
  (\du x^n + \beta^n \du t)\,,
\end{equation}
and the extrinsic curvature is given by the spatial
projection of the covariant derivative of the time like
unit normal $n_{\alpha}$ of the foliation, $K_{\alpha\beta}=-(\delta^{\mu}{}_{\beta}+n^{\mu}n_{\beta}) \nabla_{\mu} n_{\alpha}$. 
In analogy to the extrinsic curvature, we define a time
derivative for the scalar field\footnote{Equation \eqref{eq:Pi_scalar_field} is given in the conventions of the {\sc Lean} code, whilst in {\sc GRChombo} the time derivative for the scalar field is defined via $\Pi_{\rm{GRChombo}} = 2 \Pi$.}
\begin{equation} \label{eq:Pi_scalar_field}
  \Pi \defeq -\frac{1}{2\alpha} (\partial_t \varphi
  -\beta^m \partial_m \varphi)\, ,
\end{equation}
so that the energy density, momentum density and
stress-tensor can be written as
\begin{eqnarray}
  \rho &=& 2\Pi \bar{\Pi} + \frac{1}{2} \partial^m \bar{\varphi}
  \,\partial_m \varphi + \frac{1}{2}V\,,
  \nonumber \\
  j_i &=& \bar{\Pi} \partial_i \varphi
  + \Pi \partial_i \bar{\varphi}\,,
  \nonumber \\
  S_{ij} &=& \partial_{(i} \bar{\varphi}
  \partial_{j)}\varphi
  -\frac{1}{2}\gamma_{ij}
  (\gamma^{mn}\partial_m \bar{\varphi}
  \,\partial_n \varphi
  -4 \bar{\Pi}\,\Pi + V)\,.
  \label{eq:mattersources}
\end{eqnarray}
These sources appear on the right-hand side of the
Einstein equations as given in full detail in
Eqs.~(8)-(12) of Ref.~\cite{Helfer:2021brt},
and the time evolution of the scalar field is given
by the first-order-in-time system of equations
(15) and (17) in Ref.~\cite{Helfer:2021brt}. Besides the evolution equations, Einstein's equations imply the Hamiltonian, $\mathcal{H}$, and momentum, $\mathcal{M}_i$, constraints given by
\begin{align}
    \mathcal{H} &\defeq \mathcal{R} + K^2 - K^{mn}K_{mn} - 16\pi \rho = 0, \label{eq:ham_constraints}\\
    \mathcal{M}_i &\defeq D_i K - D_m K^m_i + 8\pi j_i = 0. \label{eq:mom_constraints}
\end{align}

Both codes evolve lapse  $\alpha$ and shift $\beta^i$
according to the {\it moving puncture} gauge
\cite{Campanelli:2005dd,Baker:2005vv}, i.e.~using 1+log
slicing and the $\Gamma$ driver condition as given
in Eq.~(18) of Ref.~\cite{Helfer:2021brt}. The equations are implemented in the form of finite differencing; more specifically,
we use fourth-order spatial differencing with a fourth-order
Runge-Kutta {\it method of lines}
integration in time \cite{Press1992}. Other key ingredients of the codes differ in evolving the Einstein equations, so we summarise them in the following sections and list the grid setups employed for our runs.

\subsection{Lean}
\label{sec:leancode}
The {\sc lean} code is based on the {\sc cactus} computational toolkit
\cite{Allen:1999} and employs mesh refinement in the form of moving boxes as
provided by {\sc carpet} \cite{Schnetter:2003rb}. The code evolves the Einstein equations using the Baumgarte-Shapiro-Shibata-Nakamura-Oohara-Kojima (BSSNOK) formulation
\cite{Baumgarte:1998te,Shibata:1995we,Nakamura:1987zz}, i.e.~describes
the spacetime in terms of conformally rescaled and trace-split
variables
\begin{eqnarray}
  \chi \defeq (\det \gamma_{ij})^{-1/3}\,,~~~~~~~~~~
  &
  K=\gamma^{mn}K_{mn}\,,
  &
  \nonumber \\
  \tilde{\gamma}_{ij} \defeq \chi \gamma_{ij}\,,
  &
  \tilde{A}_{ij} \defeq \chi \left(
  K_{ij}-\frac{1}{3}\gamma_{ij}K
  \right)\,,
  \nonumber \\
  \tilde{\Gamma}^i \defeq \tilde{\gamma}^{mn}
  \tilde{\Gamma}^i_{mn}\,,
\end{eqnarray}
where $\tilde{\Gamma}^i_{mn}$ are the Christoffel symbols
associated with the conformal metric $\tilde{\gamma}_{ij}$.
Apparent horizons are computed using Thornburg's {\sc ahfinderdirect} \cite{Thornburg:1995cp,Thornburg:2003sf}.

The computational domain for all {\sc lean} simulations
consists of 7 nested refinement levels: 4 outer levels centered
on the origin and 3 inner levels, each consisting of 2 boxes
centered on the two BSs. The box (edge) size decreases from
each outer level inwards by a factor of 2 except for level 4 to 5
where it decreases by a factor of 8. The grid spacing
$\du x$ decreases by a factor 2 on each consecutive inner level.
We can thus describe a grid in terms of two numbers, the edge $L_1$
and grid spacing $\du x_1$ of the outermost level. For
the {\sc lean} simulations of this paper we use\footnote{We use approximate equality here, as the total mass in our simulations is roughly one but not exactly.} $L_1 \approx 1024\,M$, $\du x_1 \approx 2.67\,M$ which implies boxes of size $L_7 \approx 4\,M$ with
spacing $\du x_7 \approx M/24$ on the innermost level.
\subsection{GRChombo}
\label{sec:grchombo}

The {\sc GRChombo} code is built on the {\sc Chombo} \cite{chombo} adaptive mesh refinement (AMR) libraries and evolves the Einstein equations using the covariant and conformal Z4 (CCZ4) formulation \cite{Alic:2011gg}. The full Einstein equations in the CCZ4 formulation can be found in Section III F of \cite{Alic:2013xsa}, where we choose $\kappa_1 \to \kappa_1/\alpha$, $\kappa_1 = 0.1$, $\kappa_2 = 0$ and $\kappa_3 = 1$. We set up a grid of length $L_1 \approx 512\,M$ with 7 additional AMR levels such that on the finest level, the spatial grid spacing is $\du x_7 \approx M/32$. Finally, we use a tagging criterion based on second derivatives of the complex scalar field and the conformal factor (see Section 3.5 of \cite{Radia:2021smk} for more details).

\subsection{Convergence testing}
The full details of the convergence testing are provided in \ref{sec:appendix_convergence}. In summary, the total error budget, including finite radius extraction and discretisation errors, is 2 \% for {\sc Lean} and 3.7 \% for {\sc GRChomobo}. All of the results reported here use extraction radius $R^{\rm{Lean}}_{\rm{ex}} \approx 200\,M$ for {\sc Lean} and $R^{\rm{GRChombo}}_{\rm{ex}} \approx 120\,M$ for {\sc GRChombo}.
\section{Boson-star binary initial data construction} \label{sec:initial-data}
The construction
of binary initial data is a challenging task in GR, mainly due to the non-linearity of the Einstein equations and the gauge dependence of the variables describing the spacetime. For boson stars, we encounter two additional challenges not present for neutron stars or black holes: (i) exponentially growing modes of single-star solutions and (ii) the lack of a consistent framework for binary initial data that are conformally flat; cf.~for example the great simplification
afforded by Bowen-York data \cite{Bowen:1980yu,Brandt:1997tf}.
In this section, we present an ansatz for computing BS-binary
initial data that significantly reduce constraint violations relative to the superposition methods used in the literature, and that 
we also expect to be a valuable preconditioner reducing
unphysical features in a full constraint solving process.

\subsection{Revisiting the initial data construction for equal-mass binaries} \label{methods_equal}

Before describing our proposed methodology, we briefly outline the equal-mass initial data construction proposed in \cite{Helfer:2018vtq}, which our method then generalises to the unequal mass case.
In the remainder of the paper, the initial set-up of our BS binary configurations is as follows. We start off with two BS stars, star A and star B, initially located at $x_{\rm{A}}^i$ and $x_{\rm{B}}^i$, and therefore separated by an initial distance $d = || x_{\rm{A}}^i - x_{\rm{B}}^i||$. We then boost the stars through Lorentz transformations with initial velocities $v_{\rm{A}}^{i}$ and $v_{\rm{B}}^{i}$ towards each other. The details of the $3+1$ variables with the Lorentz boost can be found in Ref. \cite{Helfer:2021brt}. We note that the initial positions $x_{\rm{A}}^i$, $x_{\rm{B}}^i$ and initial boost velocities $v_{\rm{A}}^i$, $v_{\rm{B}}^i$ are chosen such that the BSs are initially located in the centre of mass frame; these values will be given in our specification of the simulations in Table \ref{configurations}.

The most common procedure for constructing binary initial data is to superpose individual star solutions in a point-wise fashion. In terms of the $3+1$ ADM variables this is written as
\begin{align}
\gamma_{ij} &= \gamma_{ij}^{\rm{A}} + \gamma_{ij}^{\rm{B}} - \delta_{ij}, \label{eq:plain_sup} \\
\varphi &= \varphi_{\rm{A}} + \varphi_{\rm{B}} ,\\
\Pi &= \Pi_{\rm{A}} + \Pi_{\rm{B}}, \\
K_{ij} &= \gamma_{m(i} \left[K_{j)n}^{\rm{A}} \gamma_{\rm{A}}^{mn} + K_{j)n}^{\rm{B}} \gamma_{\rm{B}}^{nm} \right].
\label{eq:equalmassfix}
\end{align}
This method and, in particular, equation \eqref{eq:plain_sup} will henceforth be referred to as the method of \textit{plain superposition}; here the value of $\delta_{ij}$ is subtracted from the two superposed individual metrics to ensure the Minkowski metric is recovered in the far-field limit. Whilst the asymptotic flatness condition is thus satisfied, it has been shown in Refs.~\cite{Helfer:2018vtq, Helfer:2021brt} that plain superposition can induce large deviations from the equilibrium values of the volume element, $\sqrt{\rm{det}(\gamma)}$, at the centres of the stars. This effect arises from the fact that our binary system no longer contains isolated stars and simply superposing metric solutions induces a change in the volume element near the center of each BS due to the influence of its companion (see Appendix A1 of Ref. \cite{Helfer:2018vtq} for more details). To account for such a change in the volume element for the equal-mass binary stars, Ref. \cite{Helfer:2018vtq} proposes to modify the plain superposition by replacing \eqref{eq:plain_sup} with
\begin{equation} \label{eq:thomas_tick}
    \gamma_{ij} = \gamma_{ij}^{\rm{A}} + \gamma_{ij}^{\rm{B}} - \gamma_{ij}^{\rm{B}}(x_{\rm{A}}^i) = \gamma_{ij}^{\rm{A}} + \gamma_{ij}^{\rm{B}} - \gamma_{ij}^{\rm{A}}(x_{\rm{B}}^i).
\end{equation}
This modification recovers the equilibrium volume element at the centres of stars A and B as in the case of isolated stars and from now on we will refer to equation \eqref{eq:thomas_tick} as the \textit{equal-mass fix}. The equal-mass fix has been shown to significantly reduce constraint violations at the centres of the stars relative to the plain superposition procedure and also mitigate spurious physical features such as premature collapse of the stars to a BH and/or altered GW signals \cite{Helfer:2021brt}. We stress, however, that \eqref{eq:thomas_tick} is only applicable for equal-mass binaries, since in that case $\gamma_{ij}^{\rm{B}}(x_{\rm{A}}^i) = \gamma_{ij}^{\rm{A}}(x_{\rm{B}}^i)$. In the unequal-mass case, this no longer holds true: the volume element change invoked by each star on its companion will no longer be the same for both stars.

\subsection{Construction of the generalised unequal-mass initial data} \label{methods_unequal}
Now, the question arises how to generalise the proposed modification \eqref{eq:thomas_tick} to unequal-mass BS binaries. In the spirit of Eq.~\eqref{eq:thomas_tick}, we have to meet only two conditions: fix the volume element at the centres of each of the stars, $x_{\rm{A}}^i$ and $x_{\rm{B}}^i$.
One way to do so is to work with the 3-metric components $\gamma_{ij}$ directly and introduce spatially varying corrections that recover the required volume element at the centres of the stars. However, this leads to an under-determined problem, since each metric has 6 components, whilst we have only 2 conditions to satisfy. Instead, we choose to work with the conformal factor, which is a scalar density and thus makes it possible to satisfy both conditions at $x_{\rm{A}}^i$ and $x_{\rm{B}}^i$ using only 2 parameters.

We start with the plainly superposed metric \eqref{eq:plain_sup} and conformally decompose it with a conformal factor $\lambda$ defined by
\begin{equation} \label{eq:conf_metric}
  \tilde{\gamma}_{ij} = \lambda^{-1} \gamma_{ij}~~~~~\text{with}~~~~~
  \lambda = \gamma^{1/3},
\end{equation}
where $\det \tilde{\gamma}_{ij}=1$ by construction. We note that the conformal factor $\lambda$ is related to the standard BSSN/CCZ4 variable $\chi$ by $\lambda^{-1} = \chi$. \ref{sec:conf_appendix} discusses more general choices for the conformal factor and illustrates why the exponent of $1/3$ in Eq. \eqref{eq:conf_metric} is a particularly convenient choice. In our procedure, we leave the conformally rescaled metric $\tilde{\gamma}_{ij}$ unchanged at the values it takes on in the plain superposition. So far this construction does not remedy the main error inherited from the plain superposition consisting in the change of the volume element at the centres of the stars and the resulting perturbation in their central energy densities. However, we can control the volume element by adjusting the superposed conformal factor $\lambda$ with a correction $\delta \lambda$, such that we recover the equilibrium volume element at both stars' centres. At the centers of the stars this correction must satisfy 
\begin{align}
  \lambda_{\rm new}(x_{\rm A}^i) &=
  \lambda(x_{\rm A}^i) + \delta\lambda(x_{\rm A}^i) =
  \lambda_{\rm A}(x_{\rm A}^i), \label{eq:conditionAB_1} \\
 \lambda_{\rm new}(x_{\rm B}^i) &=
  \lambda(x_{\rm B}^i) + \delta\lambda(x_{\rm B}^i) =
  \lambda_{\rm B}(x_{\rm B}^i)\,,
  \label{eq:conditionAB_2}
\end{align}
where $\lambda_A$ and $\lambda_B$ are the unperturbed conformal factors for stars $A$ and $B$.
In contrast to the equal-mass case, this approach necessitates a spatially varying correction to the conformal factor. For this purpose, we construct weight functions $w_{\rm A}(x^i)$ and $w_{\rm B}(x^i)$ around the centres of the stars, which will be specified later in this Section. We then propose the following ansatz for the new conformal factor on the entire initial hypersurface

\begin{equation}
     \lambda_{\rm new}(x^i) =
  \lambda(x^i) + w_{\rm A}(x^i) h_{\rm A} + w_{\rm B}(x^i)h_{\rm B}\,.
  \label{eq:lambdanew}
\end{equation}
Here $h_{\rm A}$ and $h_{\rm B}$ are determined by imposing our target conditions \eqref{eq:conditionAB_1}--\eqref{eq:conditionAB_2}, which reduce to a $(2\times 2)$ system of linear equations. The required values of $h_{\rm A}$ and $h_{\rm B}$ are then found to be
\begin{equation}
  h_{\rm A} =
  \frac{-w_{\rm B}(x_{\rm{A}}^i)\delta \lambda(x_{\rm{B}}^i)
        + w_{\rm B} (x_{\rm{B}}^i)\delta \lambda (x_{\rm{A}}^i)}
       {w_{\rm A}(x_{\rm{A}}^i) w_{\rm B} (x_{\rm{B}}^i) - w_{\rm A}(x_{\rm{B}}^i)w_{\rm B}(x_{\rm{A}}^i)}
       \,,~~~~~
  h_{\rm B} =
  \frac{w_{\rm A}(x_{\rm{A}}^i)\delta \lambda(x_{\rm{B}}^i)
        - w_{\rm A}(x_{\rm{B}}^i)\delta \lambda (x_{\rm{A}}^i)}
       {w_{\rm A}(x_{\rm{A}}^i) w_{\rm B}(x_{\rm{B}}^i) - w_{\rm A}(x_{\rm{B}}^i)w_{\rm B}(x_{\rm{A}}^i)}\,.
  \label{eq:hAB}
\end{equation}
For these constants $h_{\rm{A}}$ and $h_{\rm{B}}$ the newly corrected conformal factor $\lambda_{\rm{new}}$ allows us to recover the desired volume element at the centres of the stars, as if they were isolated. This can be seen by considering the newly re-scaled metric
\begin{equation} \label{eq:gamma_new}
    \gamma^{\rm{new}}_{ij} = \left(\frac{\lambda_{\rm{new}}}{\lambda} \right) \gamma_{ij} = \frac{\lambda_{\rm{new}}}{\gamma^{1/3}} \gamma_{ij},
\end{equation}
where $\gamma^{-1/3} \gamma_{ij}$ has unit determinant by construction and therefore ensures that $\gamma^{\text{new}} (x_{\rm A}^i) = \gamma^{\rm A}(x_{\rm A}^i)$ and likewise for star B.

We are now left to choose what weight functions to use around the stars in Eq. \eqref{eq:lambdanew}. Focusing here on asymptotically flat spacetimes, we wish to obtain metric corrections that fall off $\propto 1/r$. To guarantee such behaviour we construct the following weight functions
\begin{align} \label{eq:weight_function}
  w_{\rm J}(x^i) &= \frac{1}{\sqrt{R_{\rm J}^2 + r_{\rm J}^2}}
\end{align}
where $\rm{J} \in \{\rm{A},\rm{B} \}$, $r_{\rm{J}}\defeq ||x^i-x^i_{\rm{J}}||$ and $R_{\rm{J}}$ are freely specifiable constants that control the width of the functions. 
Our initial data method therefore consists of the following main steps:
\begin{enumerate}
    \item Construct the plainly superposed metric $\gamma_{ij}$ according to \eqref{eq:plain_sup}.
    \item Construct the conformal factor $\lambda$ from a plainly superposed metric according to \eqref{eq:conf_metric}.
    \item Choose a suitable parameter pair $(R_{\rm{A}}, R_{\rm{B}})$.
    \item Compute corrections at the stars' centres, $\delta \lambda (x_{\rm A})$ and $\delta \lambda (x_{\rm B})$, according to Eqs.~\eqref{eq:conditionAB_1}-\eqref{eq:conditionAB_2} required for calculation of constant $h_{\rm A}$ and $h_{\rm B}$ in Eq.~\eqref{eq:hAB}.
    \item Correct the conformal factor $\lambda$ to $\lambda_{\rm{new}}$ in Eq.~\eqref{eq:lambdanew} to recover the equilibrium volume element at the stars' centres.
\end{enumerate}
The choice of the two parameters, $R_{\rm{A}}$ and $R_{\rm{B}}$, in step (ii) will be explored in more detail through our numerical simulations in the next Section. 

\section{Set-up and exploration of the parameter space} \label{sec:parameter_space}
In this section, we start by listing the BS models we study in this work in Table \ref{models}. We focus on the solitonic BSs, with $\sqrt{G}\sigma_0=0.2$, which allows us to access a variety of mass ratios, including heavy and relatively compact BSs. To construct binary systems of various mass ratios $q = M_{\rm A} / M_{\rm B}$, where $M_{\rm A} < M_{\rm B}$, we superpose combinations of the BS models from Table \ref{models}. The resulting binary configurations are detailed in Table \ref{configurations}.
\begin{table}
\centering
\footnotesize
\begin{tabular}{C{1.5cm}C{1.5cm}C{1.5cm}C{1.5cm}C{1.5cm}C{2.5cm}C{2.5cm}}
\br                              
Model&$\sqrt{G}A(0)$&$\mu M_{\rm BS}$&$\omega/\mu$ & $\mu r_{99}$ & \rm{max}($m(r)/r$) \\
\mr
\texttt{S-170} & 0.17 & 0.7134 & 0.4392 & 3.97 & 0.222 \\
\texttt{S-160} & 0.16 & 0.5368 & 0.5375 & 4.18 & 0.166 \\
\texttt{S-147} & 0.147 & 0.3606 & 0.6784 & 4.48 & 0.115  \\
\texttt{S-100} & 0.1 & 0.2701 & 0.8506 & 6.21 & 0.070 \\
\br
\end{tabular}
\caption{\label{models}Solitonic BS models with $\sqrt{G}\sigma_0=0.2$ considered in this work. $A(0)$ denotes the central scalar field amplitude, $M_{\rm BS}$ the mass of the BS, $\omega$ the frequency of the ground state solution, $r_{99}$ the areal radius containing $99 \%$ of the BS mass, and \rm{max}($m(r)/r$) our measure of compactness. Note that for this potential the maximum mass of a BS is $\mu M_{BS} = 0.7212$}. 
\end{table}

\begin{table}
\centering
\footnotesize
\begin{tabular}{C{0.8in} C{0.6in} C{0.6in} C{0.5in} C{0.5in} C{1.2in} C{0.3in} C{0.8in}}
\br                                
Run & Model for star A& Model for star B & $v_{x}^{\rm{A}}(0)$ & $v_{x}^{\rm{B}}(0)$ & $d/M$ & $q$ & Code \\ 
\mr
\texttt{q1-dX-pX} & \texttt{S-170} & \texttt{S-170} & $-0.1$ & $0.1$ & $11.2, 22.3, 33.5, 44.6$ & $1$ & both  \\
\texttt{q075-dX-pX} & \texttt{S-160} & \texttt{S-170} & $-0.1141$ & $0.0859$ & $12.7, 25.5, 38.2, 50.9$ & $0.75$ &  {\sc GRChombo} \\
\texttt{q05-dX-pX} & \texttt{S-147} & \texttt{S-170} & $-0.1328$ & $0.0672$ & $14.8, 29.7, 44.5, 59.3$ & 0.5 & \sc{Lean}  \\
\texttt{q038-dX-pX} & \texttt{S-100} & \texttt{S-170} & $-0.1451$ & $0.0549$ & $16.2, 32.4, 48.6, 64.8$ & $0.38$ & both   \\
\br
\end{tabular}
\caption{\label{configurations}Binary initial data configurations considered in this work. Each run has a suffix \texttt{dX-pX}, where \texttt{dX} is a wildcard for the initial separation and \texttt{pX} is a wildcard for the off-phase parameter in degrees, $\frac{180^{\circ}}{\pi} \delta \phi \in [0^{\circ} , 360^{\circ})$. In our convention, the off-phase parameter, $\delta \phi$, is added to star A, whilst for star B it remains zero. Here $v_x^{\rm J}$ for $\rm{J}\in\{{\rm A, B}\}$ denote the initial boost velocities with associated Lorentz factors $\gamma_{\rm J}$, $d =||x_{\rm A}^i - x_{\rm B}^i||$ is the initial separation and $M = \gamma_{\rm A} M_{\rm BS}^{\rm A} + \gamma_{\rm B} M_{\rm BS}^{\rm B}$ is the total mass of the binary.}
\end{table}

\subsection{Equal-mass binaries}
\begin{figure}[hbt!]%
    \centering
    \includegraphics[width=7.5cm,valign=c, clip=True]{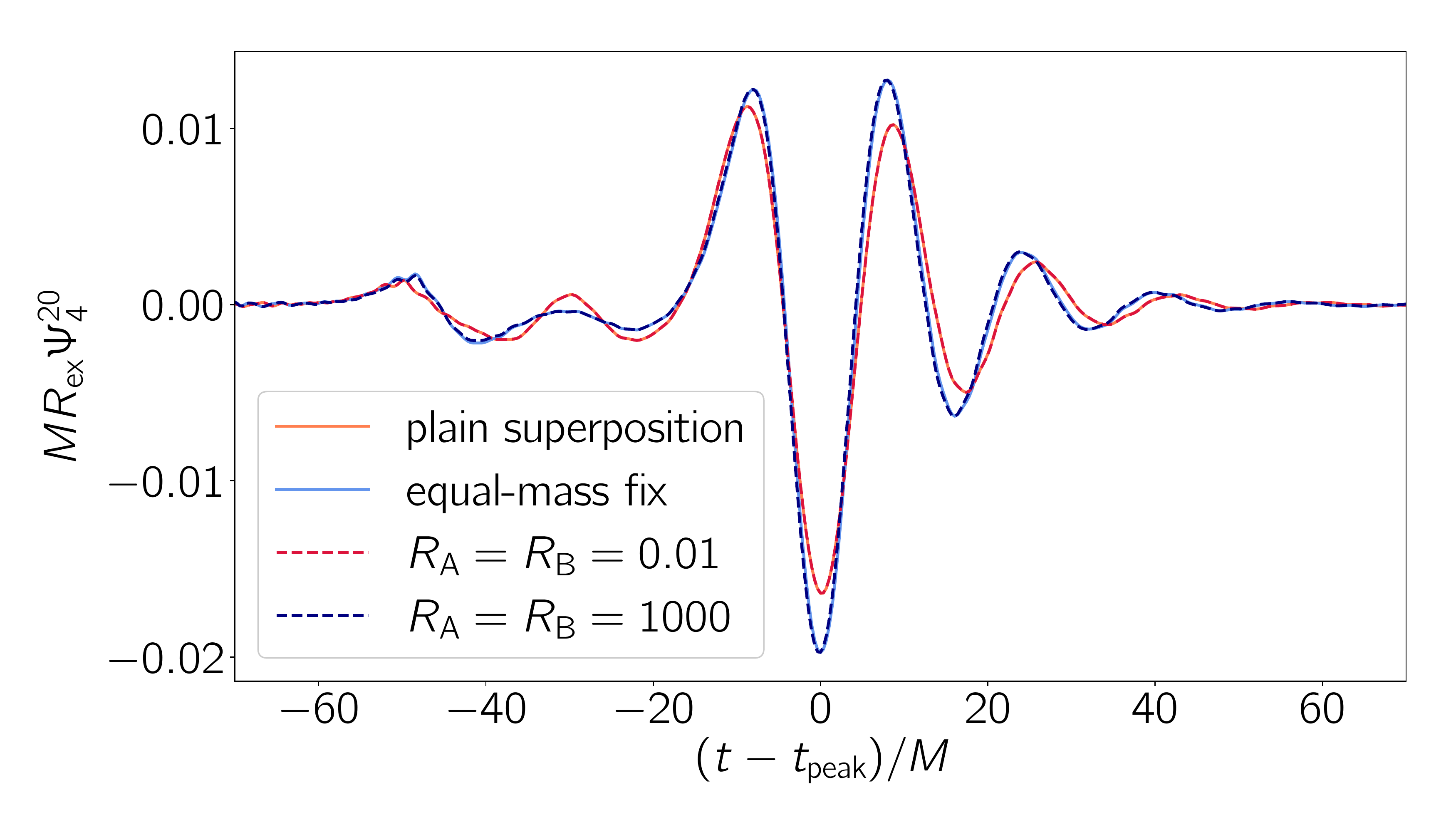}%
    \qquad
    \includegraphics[width=7.5cm,valign=c, clip=True]{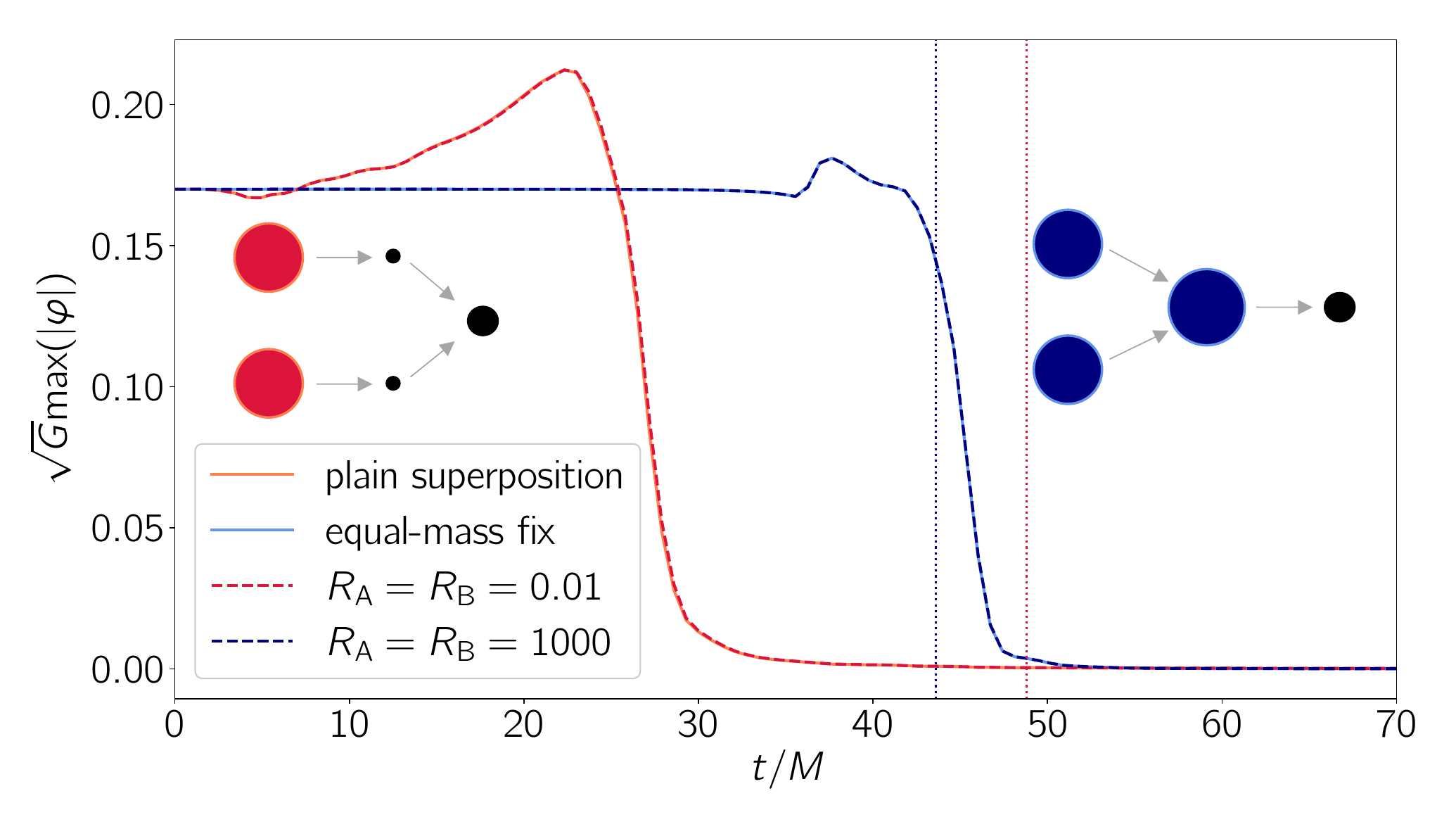}%
    \label{fig:q1_comparison}%
    \caption{\textit{Left:} Real parts of the $(20)$-mode of the Newman-Penrose scalar $\Psi_{4}$ for binary sequence \texttt{q1-d11-p000}, obtained for plain superposition, the equal-mass fix and our improved method. The waveforms have been shifted by $t_{\rm{peak}}$, the time at which the maximum GW amplitude is reached.
    \textit{Right:} Maximum of the scalar field amplitude for the same binary configuration. Notably, plain superposition forms a BH well before the equal-mass fix, therefore resulting in altered gravitational waveform. Here we indicate the merger time by the vertical lines.
    }
\end{figure}

For equal-mass binaries, the equal-mass fix \eqref{eq:thomas_tick} has been shown to remedy the spurious effects of plain superposition \cite{Helfer:2021brt}. It is therefore important to verify that our method recovers these improvements in the equal-mass limit. First, we recall that our method requires the choice of free
parameters $(R_{\rm A}, R_{\rm B})$. As shown in more detail
in \ref{sec:limit_cases}, we can recover both, plain
superposition and the equal-mass fix, as limiting cases of
this choice. Specifically, in the limit $R_{\rm A},
R_{\rm B}\rightarrow 0$, our initial data reconstruction recovers plain superposition, whilst for the equal-mass
case and in the limit $R_{\rm A},R_{\rm B}\rightarrow \infty$,
we recover exactly the equal-mass fix \eqref{eq:thomas_tick}.
Our method thus provides a direct generalization for constructing
BS binary initial data with arbitrary mass ratios that includes
plain superposition and the equal-mass fix as limiting cases.
We next verify this claim empirically by evolving in time
the binary configuration \texttt{q1-d11-p000} using four
types of superposition: plain, the equal-mass fix and our
generalized method using very small and very large
$(R_{\rm A},R_{\rm B})$, namely $R_{\rm A}=R_{\rm B}=0.1$
and $1000$. Figure \ref{fig:q1_comparison} shows the
gravitational waveforms and the central scalar field value
as functions of time obtained for these four cases.
The figure demonstrates excellent agreement of our
proposed method with plain superposition and the equal-mass
fix, respectively, for $R_{\rm A}=R_{\rm B}=0.01$ and
$R_{\rm A}=R_{\rm B}=1000$. 

\subsection{Unequal-mass binaries} \label{sec:parameter_space_qneq1}
\begin{figure}%
    \centering
    \includegraphics[width=8cm,valign=c]{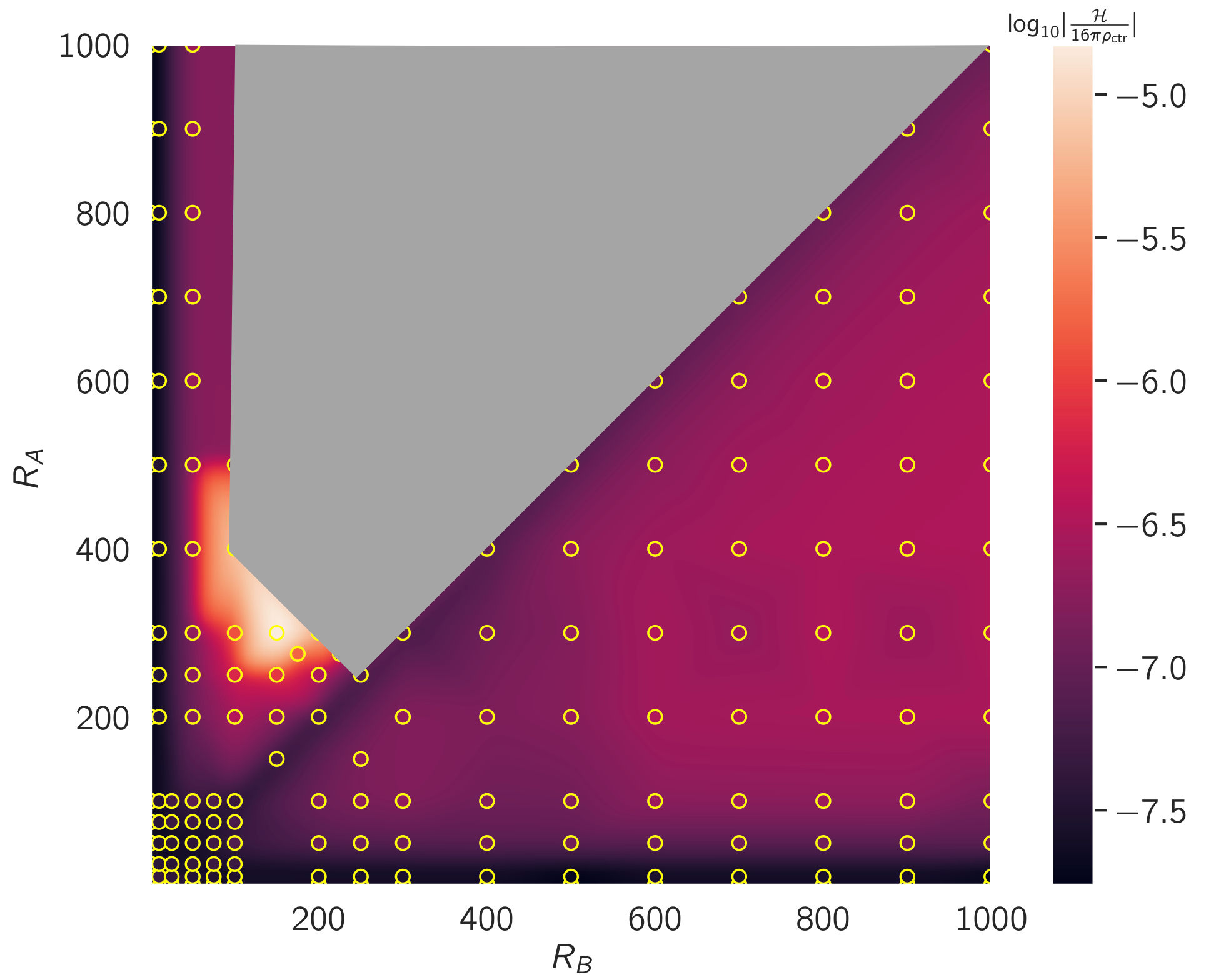}%
    \qquad
    \includegraphics[width=7.8cm,valign=c]{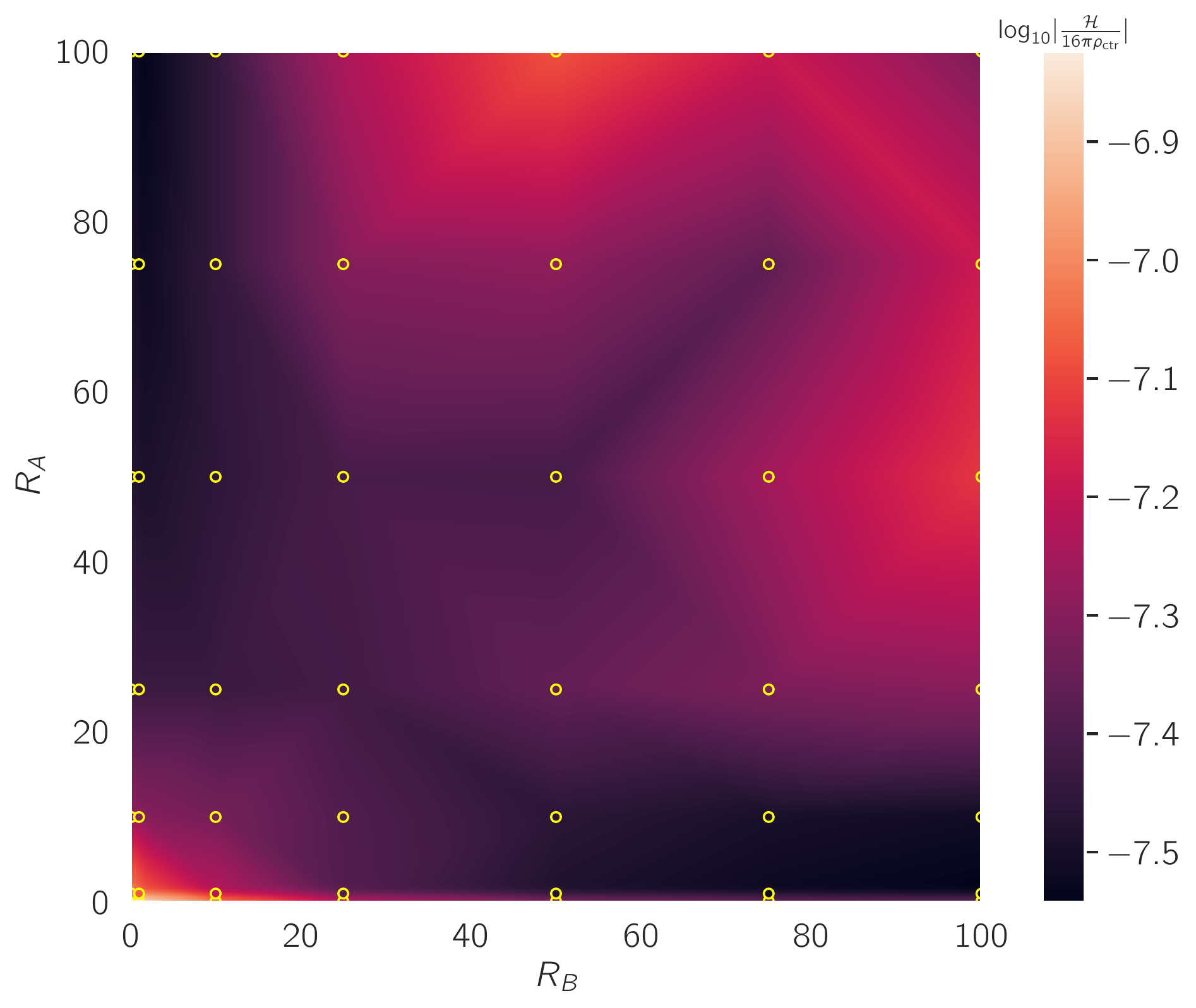}%
    \label{fig:heat_map_q075}%
    \caption{\textit{Left:} Log plot of the $L2$-norm of the Hamiltonian constraint violations (normalised by the central energy density
    $16\pi \rho_{\rm ctr}$ of the heavier star) across the simulation grid for the binary configuration \texttt{q075-d12-p000}. The grey region indicates the parameter space of $(R_{\rm{A}}, R_{\rm{B}})$ where constraint violations diverge. The values of constraint violations have been interpolated onto the full domain using the runs performed indicated by the yellow circles. \textit{Right:} A zoom-in on the region $(R_{\rm{A}}, R_{\rm{B}}) \in [(0,100) \times (0,100)]$. Asymmetries in the constraint violations are visible across the diagonal $R_{\rm{A}} = R_{\rm{B}}$, demonstrating the non-trivial dependence of the constraints on the choice of radial parameters. In the equal-mass case, we have verified that the constraint violations are symmetric under exchange ${\rm A}\leftrightarrow {\rm B}$, as expected. The results for the remaining mass ratios are displayed in Figure~\ref{fig:heatmaps_q05q038}.}
\end{figure}

In the unequal-mass case, the choice of radial parameters  $(R_{\rm{A}}, R_{\rm{B}})$ is more complex. Changing the profile shape of the radial functions affects the extent to which the corrections are applied to the conformal factor in Eq.\eqref{eq:lambdanew} around the centres of the stars: very small $(R_{\rm{A}}, R_{\rm{B}})$ will result in smaller corrections around the stars and vice versa. As such, there exists a region of suitable radial parameters, which we find numerically by calculating the $L2$-norm of the Hamiltonian constraint \eqref{eq:ham_constraints} in the parameter space of $(R_{\rm{A}}, R_{\rm{B}})$. Figure \ref{fig:heat_map_q075} demonstrates the dependence of the $L2$-norm of the Hamiltonian constraint violations on the choice of radial parameters for a binary configuration with $q=0.75$. We show the corresponding "heat-maps" for other mass ratios in \ref{sec:heat_maps_appendix}. It is important to note that there are regions in the parameter space, where the initial constraint violations become very large -- these regions are indicated by the grey shaded area for readability. The diverging behavior of the constraints for certain pairs of $(R_{\rm A}, R_{\rm B})$ can be attributed to a zero crossing of $\lambda_{\rm new}$ away from the stars' centers, which results in a singular 3-metric. Regions in the parameter space where this happens are evidently not suitable for time evolutions.

 In the following, we base our specific choices for radial parameters $(R_{\rm A}, R_{\rm B})$ on two criteria. First, the pair has to result in reduced constraint violations relative to plain superposition, and second, they should remedy spurious oscillations in the time evolutions of the scalar field profiles that result from plainly superposed initial data\footnote{Note that these two criteria are not the same; it is for instance possible to choose $(R_{\rm A}, R_{\rm B})$ that reduce the constraint violations without satisfactorily reducing scalar field oscillations.}. In this section we focus on constraint violations, whilst we discuss the scalar field profiles in more detail in Section \ref{sec:Actr_profiles}. For unequal-mass BS binaries, we find that radii $R_{\rm{A}} = 10-100$ (light BS) and $R_{\rm{B}} = 1-100$ (heavy BS) work generally well. For smaller mass ratios $q \lesssim 0.5$ we find that $(R_{\rm A}, R_{\rm B})$ differing by at most one order of magnitude are optimal. We note that different pairs $(R_{\rm{A}}, R_{\rm{B}})$ can result in small global time-shifts in the GW signals due to gauge effects; besides this time shift, however, the phase and amplitude of the waveforms remain unaffected. We summarise our choices of radial parameters in Table \ref{radial_choices} for all unequal-mass binary configurations studied here.

\begin{figure}%
    \centering
    \includegraphics[width=7.5cm,valign=c,clip=True]{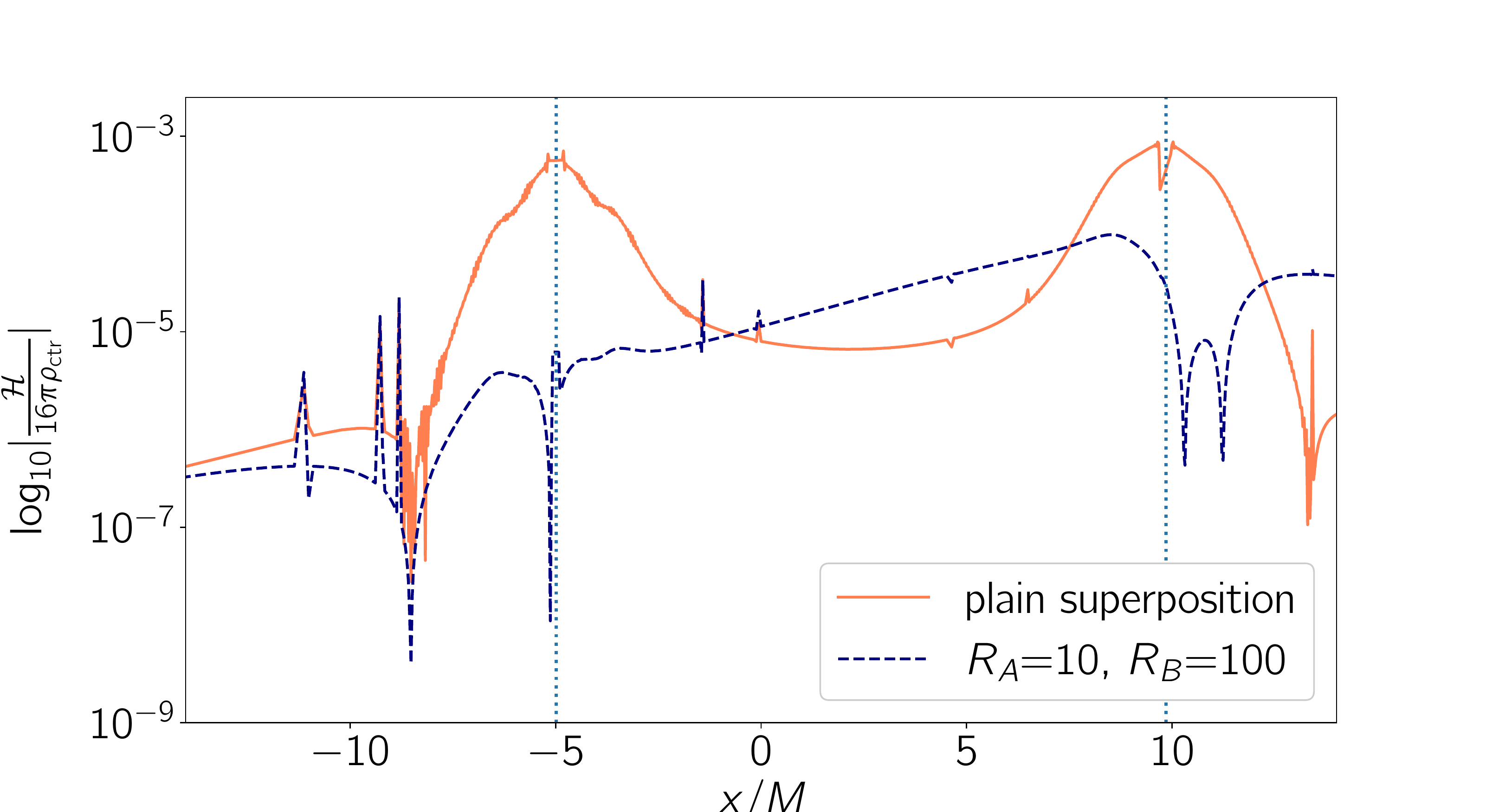}%
    \qquad
    \includegraphics[width=7.5cm,valign=c,clip=True]{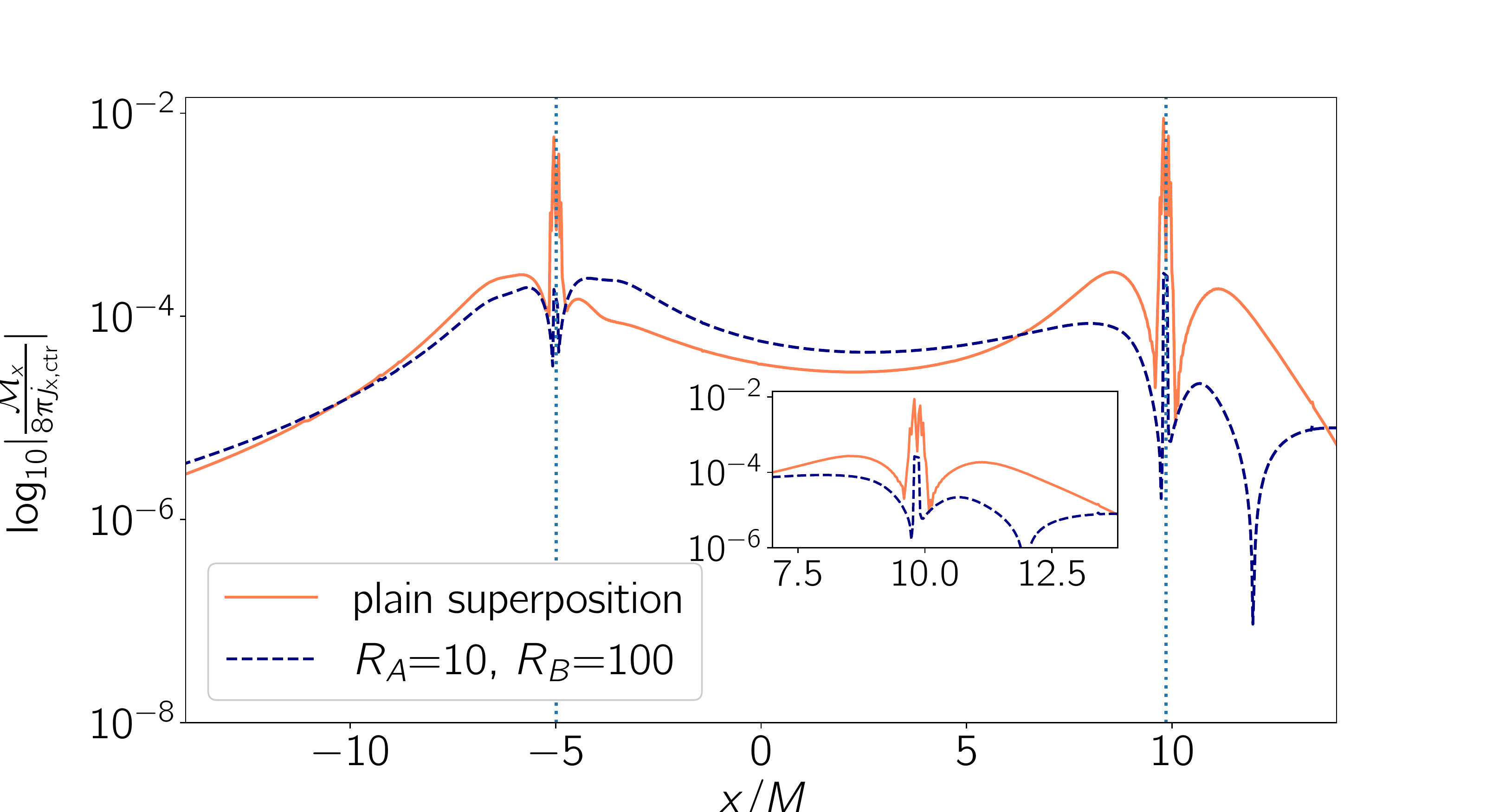}%
    \label{fig:constraints_q05}%
    \caption{Log-scaled Hamiltonian constraint $\mathcal{H}$ (normalised by the central energy density $16\pi \rho_{\rm ctr}$ of the heavier star) and the x-component of the momentum constraint $\mathcal{M}_x$ (normalised by the central momentum density $8\pi j_{x,\rm ctr}$ of the heavier star) computed along the collision axis of the binary of mass ratio $q=0.5$ and initial separation $d/M = 14.8$.}
\end{figure}

For the radial parameters thus chosen, we observe a particularly pronounced reduction in the constraint violations at the centres of the stars relative to plain superposition. This is as desired by construction of our method, since the correction to the volume element is imposed exactly at the stars' centres. As for the equal-mass fix studied in Ref.\cite{Helfer:2021brt}, we find this effect to be particularly pronounced in the Hamiltonian constraint. This is illustrated in Figure \ref{fig:constraints_q05}, which shows the constraint violations along the collision axis ($x$-axis) for the binary \texttt{q05-d15-p000}.
Other binary configurations result in qualitatively similar constraint violation profiles. 

\begin{table}
\centering
\footnotesize
\begin{tabular}{C{0.8in} C{0.5in} C{0.5in}}
\br                                
Run & $R_{\rm A}$ & $R_{\rm B}$ \\ 
\mr
\texttt{q075-dX-pX} & 100 & 1   \\
\texttt{q05-dX-pX}  & 10 & 100  \\
\texttt{q038-dX-pX}  & 10 & 100   \\
\br
\end{tabular}
\caption{\label{radial_choices}Choice of radial parameters $(R_{\rm{A}}, R_{\rm{B}})$ for the unequal-mass binaries considered in this work.}
\end{table}

\section{Results: time evolutions of off-phase BS binaries} \label{sec:results}
\subsection{Equal-mass collisions} \label{sec:equal_collisions}
\begin{figure}%
    \centering
    \includegraphics[width=0.98\textwidth]{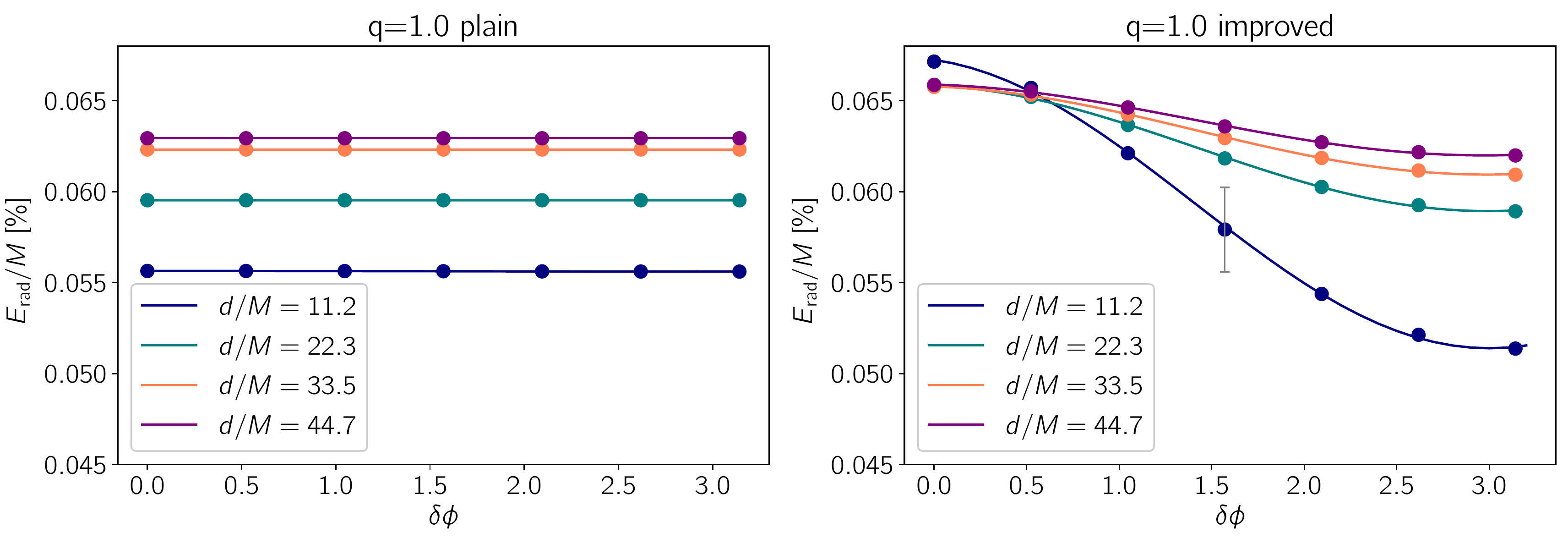}
    \caption{Radiated energy $E_{\rm rad}$ as a function of the phase off-set parameter $\delta \phi$ for equal-mass BS binary head-on collisions \texttt{q1-dX-pX} of Table \ref{configurations} starting from different
    separations $d/M$ using
    plain superposition (left panel) and
    the equal-mass fix (right panel).
    The single error bar displayed at $\delta\phi = \pi/2$ in the right panel
    indicates our numerical uncertainty which is very similar for all data points. 
    }
    \label{fig:energy_fit_equal}
\end{figure}

We begin the discussion of our results with the simplest case of equal-mass collisions simulated using plain superposition and the equal-mass fix. It has been shown in Ref. \cite{Helfer:2021brt} that for $q=1$ binaries with $\delta \phi = 0$, plain superposition results in two crucial spurious features:
\begin{enumerate}
    \item premature BH formation as indicated by a sudden drop in the scalar-field amplitude at the BS center and AH formation; i.e.~the two BSs collapse to a BH prior to merger (cf. Figure 9 of Ref. \cite{Helfer:2021brt}), which can be attributed to the spurious oscillations of the scalar field's central amplitude,
    \item energy dependence on the initial separation, which was found to be less pronounced for solitonic BSs.
\end{enumerate}
Extending the argument of Ref. \cite{Helfer:2021brt} to the case of arbitrary $\delta \phi$, we span the dephasing parameter space over the range\footnote{The range $\delta \phi \in (\pi, 2 \pi]$ is automatically covered by symmetry.} $\delta \phi \in [0, \pi]$. Similar to Ref. \cite{Helfer:2021brt}, we find that plain superposition results in premature BH formation regardless of the dephasing parameter $\delta \phi$. Therefore, in case of plain superposition, the radiated energy becomes independent of the dephasing angle, 
as is demonstrated in the left panel of Figure \ref{fig:energy_fit_equal}. In contrast, the equal-mass fix avoids premature BH formation
for all choices of $\delta \phi$ and the radiated energy takes on a non-trivial dependence on the dephasing parameter as displayed in the right
panel of Fig.~\ref{fig:energy_fit_equal}. This panel furthermore demonstrates that for large $\delta \phi$ the second observation in the
above list no longer holds: for $\delta \phi \gtrsim 1$, the radiated energy
{\it does} vary significantly with initial separation. 
As we will discuss in more detail below, the choice of the dephasing
parameter can significantly affect the dynamics and GW energy emission
for equal as well as for unequal-mass BS binaries.

The role of the phase offset has been studied before in the context of head-on collisions of Proca stars and was found to significantly affect the GW emission and mode structure of the stars \cite{Sanchis-Gual:2022mkk}. Similarly, in Ref.~\cite{Widdicombe:2019woy} the phase parameter was found to impact the merger dynamics of real-scalar-field solitons, aka oscillatons (OSs).
In particular, over
a considerable range of compactness values they find anti-phase ($\delta \phi = \pi$) OS head-on collisions to bounce whereas equal-phase ($\delta \phi = 0$) collisions
result in dispersal of the scalar field; cf.~their Fig.~1.
A similar repulsive effect has been found for $\delta \phi=\pi$ in BS
head-on collisions in Ref.~\cite{Palenzuela:2006wp, Bezares:2017mzk}, where in particular Appendix B of Ref.~\cite{Palenzuela:2006wp} gives an explanation in terms of an effective interaction potential.
This feature may be connected to the fact that anti-phase collisions produce destructive interference, as shown in the case of Newtonian gravity in Ref. \cite{Schwabe:2016rze}. Our collision sequences $\texttt{q1-dX-pX}$ involve highly compact solitonic stars, and so a BH forms post-collision. We therefore do not observe bounces in the anti-phase collisions, but the scalar field's repulsive character still manifests itself in a weaker signal and reduced radiated energy. Equal-phase configurations form a BH most 'efficiently' and result in the largest energy burst. 

The energy dependence on the phase off-set is most naturally modelled as a single sinusoidal function\footnote{Note that in the equal-mass collisions the two stars oscillate at the same frequency. For $\delta \phi = 0$ they therefore have identical phases throughout the entire infall. By perturbing the phase of one of the stars with $\delta \phi$, we change the phase difference at merger $\delta \phi_{\rm{merger}}$ by the same amount, i.e. $\delta \phi_{\rm{merger}} = \delta \phi$.}; cf. Eq.~(16) in Ref.~\cite{Schwabe:2016rze}. This is confirmed by our numerical results in Figure \ref{fig:energy_fit_equal}, which are well fitted by
\begin{equation} \label{eq:sinfit_1}
    E_{\text{fit}} = A_1 \text{sin}(f_1 \delta \phi + p_1) + s,
\end{equation}
where amplitude $A_1$, frequency $f_1$, phase $p_1$ and shift $s$ are determined using a least-squares algorithm.
From the right panel of Figure \ref{fig:energy_fit_equal}, it is clear that in the evolutions starting from the equal-mass fix, some energy discrepancy occurs between various separations.
Whereas for $\delta \phi \lesssim 1$ the radiated energy varies
only mildly with $d/M$,
it increases significantly with initial
separation for larger dephasing parameters. 
In all cases, however,
we observe a gradual convergence of the energy
for large $d/M$, albeit at distinct rates for different $\delta \phi$. 
These effects can be attributed to the increase in the collision velocity that results from larger separations and enhances the merger dynamics.
We can likewise attribute the decrease of $E_{\rm rad}$ for
larger $\delta \phi$ to the off-phase scalar fields' repulsion
and a consequential weakening of the merger dynamics. However, a higher collision velocity (equivalent to larger initial separation) appears to 'break down' the repellent nature of the scalar field, thus narrowing the energy discrepancy as $d \to \infty$. For the collisions starting from plain superposition,
we also observe convergence of the radiated energy in the limit
of large separations, but here the energy is a constant function
of $\delta \phi$ for each given $d$.

The increase of energy with separation due to higher collision velocity is a plausible interpretation, supporting our results.
But is it correct?
We quantitatively test the hypothesis as follows. Using the Newtonian approximation, we estimate that in the evolution starting from initial separation $d/M = 22.3$ and initial boost velocity $v = 0.1$, a velocity $v = 0.18$ is reached at distance $d/M = 11.2$. BS collisions with $d/M=11.2$ and this larger velocity
$v=0.18$ would then be expected to result in energy values
comparable to those obtained for the $d/M=22.3$, $v=0.1$ sequence.
Likewise, BS collisions starting with $d/M=11.2$ and $v=0.199$ $(0.208)$ should reproduce the coral (purple) curves for
$d/M=33.5$ $(d/M=44.7)$ in Fig.~\ref{fig:energy_fit_equal}. 

\begin{figure}
  \centering
  \includegraphics[width=\textwidth]{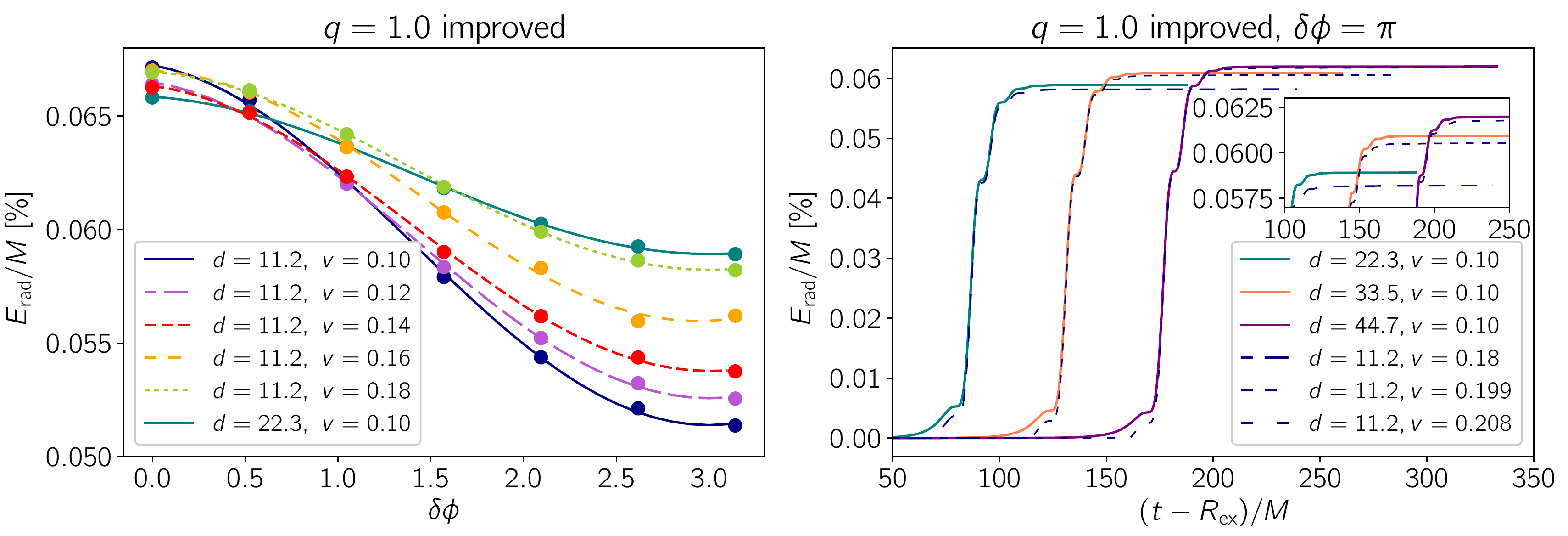}
  \label{fig:transition}
  \caption{{\it Left panel:} The radiated energy $E_{\rm rad}$ is
  shown as a function of the phase offset parameter $\delta\phi$
  as in Fig.~\ref{fig:energy_fit_equal}, but now for BS binary
  configurations starting from initial separation $d/M=11.2$
  with different velocities. For $v=0.18$, the radiated
  energy closely matches the teal colored curve of the $d/M=22.3$,
  $v=0.1$ binary as predicted by a Newtonian calculation of the
  infall velocity. {\it Right panel}: The radiated energy is shown
  as a function of time for BS binaries with a phase offset $\delta \phi=\pi$.
  Again, starting a BS binary from $d/M=11.2$ with a larger velocity,
  as obtained from a Newtonian calculation, the energy functions
  for $v=0.1$ but larger initial separation are recovered.
  The initial separation $d$ is given in units of $M$, even though
  this factor has been omitted in the legends for presentation purposes. 
  }
\end{figure}

In simple terms, we should obtain a transition of the dark blue
curve for $d/M=11.2$ in Fig.~\ref{fig:energy_fit_equal} into
the teal colored one for $d/M=22.3$ by fixing $d/M=11.2$
and increasing the initial
velocity from $v=0.1$ to $v=0.18$.
The left panel of Figure \ref{fig:transition} illustrates this transition by displaying the energy obtained for equal-mass collisions with initial separation $d/M = 11.2$ and varying initial boost velocity in the range $v \in [0.1, 0.18]$.  For $\delta \phi \gtrsim 1$ we observe a gradual increase in the energy with rising initial boost velocity.
This increase is most pronounced for the anti-phase sequence
whereas the velocity change has almost no effect on the radiated energy for in-phase binaries. Over the entire range
$\delta \phi \in [0,\pi]$, however, we recover with high accuracy
the radiated energies
of the $d/M=22.3$ binaries by starting from $d/M=11.2$ with larger
velocity $v=0.18$ as predicted by the Newtonian calculation. We are likewise able to recover comparable energies for separations of $d/M = 33.5$ and $d/M = 44.7$ by starting the sequence with $d/M = 11.2$ with yet higher initial boost velocities; see the right panel of Figure \ref{fig:transition}.

\subsection{Unequal-mass collisions} \label{sec:results_unequal}
In the remainder of this section, we focus on the unequal-mass binary simulations and present a direct comparison between the results obtained for plain superposition \eqref{eq:plain_sup} and our improved method for $q \neq 1$ binaries developed in Section \ref{methods_unequal}. There are two limits, in which we would expect the two methods to give comparable results:
\begin{enumerate}
    \item With decreasing mass ratio, the metric of the lighter star will approach Minkowski, hence the volume factor change induced by it on the heavier star will be negligible.
    For plainly superposed data, overall constraint violations will be reduced, however, the volume factor change induced on the lighter star by its heavier companion would be inevitable, resulting in spurious star excitations.
    \item In the limit of infinite separation ($d \to \infty$), both stars will be isolated and therefore even start in their 'equilibrium' state for plain superposition.
\end{enumerate}
In practice, we are limited to finite mass ratios and initial separations, and as we will see later, the regimes, where plain superposition can give comparable results to our method require very large initial separations, often impractical due to the ensuing
computational costs.

For this comparison, it is important to realize that in the unequal-mass case, the initial dephasing parameter $\delta \phi$ no longer represents the dephasing at merger,  $\delta \phi_{\rm{merger}}$.
This is 
a consequence of the two stars' different oscillation frequencies
which introduce a ``natural'' time-dependent phase offset. Introducing a constant phase offset parameter $\delta \phi$ to one of the stars adds a constant phase difference to this time dependent
dephasing in a controlled manner.
Using multiple values of the initial dephasing parameter $\delta \phi$ for otherwise
identical configurations allows us to cover a complete range of dephasing at merger, $\delta \phi_{\rm merger} \in [0, 2\pi)$. 

As will be shown in the following analysis, the key effects of varying the off-set parameter $\delta \phi$ for a given BS binary configuration are as follows: 
\begin{enumerate}
    \item a change in the infall time (Figure \ref{fig:psi4_q038_deltaphi}, $\delta \phi = 60^{\circ},150^{\circ}$), 
    \item a change in the GW amplitude and the radiated energy (Figures \ref{fig:psi4_q075_q038} and \ref{fig:energy_fit}),
    \item a relative enhancement of higher-order multipoles in the GW signal, indicating significant tidal deformation of the lighter star (Figure \ref{fig:psi4_q038_deltaphi}, $\delta \phi = 60^{\circ},90^{\circ}$). 
\end{enumerate}
First, however, we test
our improved initial data construction by exploring the
time evolution of the BSs' central scalar field amplitude. 

\subsection{Scalar field profiles} \label{sec:Actr_profiles}
The deficiencies of the plain-superposition procedure
are diagnosed most directly in the time evolution of
the scalar-field amplitude at the centres of the two BSs.
For equal mass head-on collisions of BSs and OSs,
respectively, this has been shown in Fig.~9 of
Ref.~\cite{Helfer:2021brt} and Fig.~7 of Ref.~\cite{Helfer:2018vtq}. A closer analysis of the scalar-field evolutions we
obtain from our equal-mass BS collisions of Sec.~\ref{sec:equal_collisions} confirms this picture for all
values of the dephasing parameter $\delta \phi$. In this section, we demonstrate
that for sufficiently compact BSs plain superposition results in the same spurious
effects as in the unequal-mass BS collisions, namely spurious
oscillations of the two stars' central scalar amplitudes around their equilibrium values that subsequently trigger
premature collapse to a BH.

\begin{figure}%
    \centering
    \includegraphics[width=\textwidth]{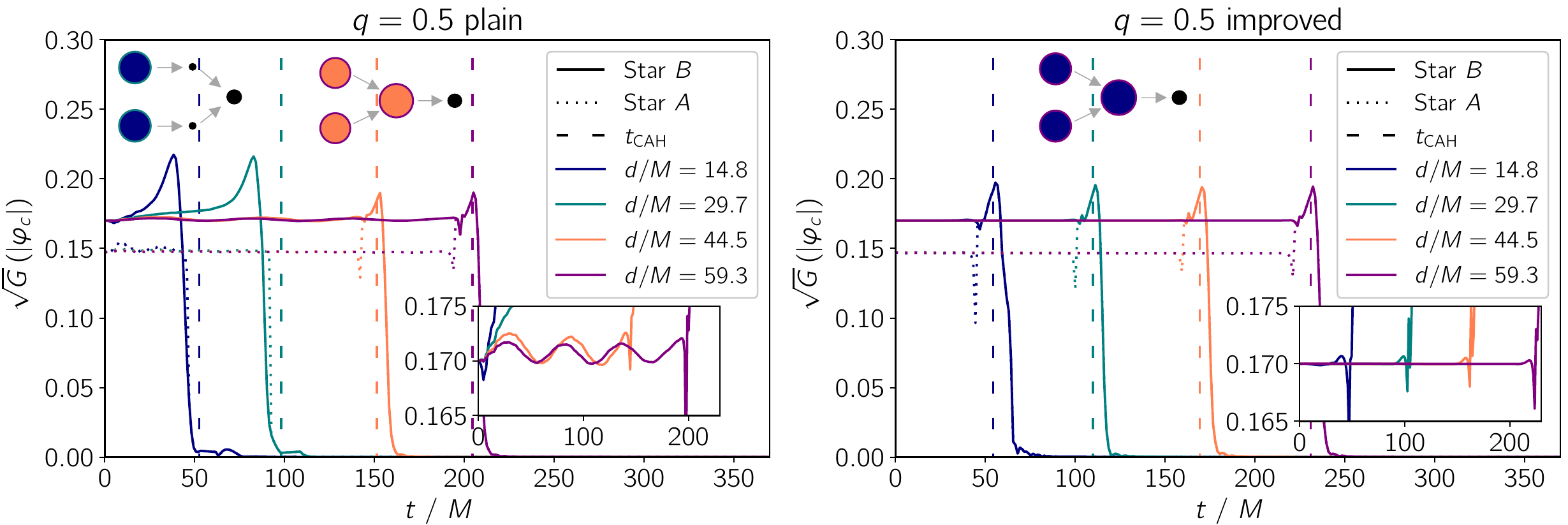}
    \label{fig:phi_q05}%
    \caption{
    Evolution of the central scalar field $|\varphi_c|$
    of the two BSs for the mass ratio $q=0.5$ using
    plain superposition (left) and the improved method
    (right). The results for the heavy star B are shown as
    solid and those for the light star A as dotted curves,
    and the color encodes the initial separation. For each
    configuration, the phase offset $\delta \phi$ has been chosen to maximize the GW radiation; cf.~Table \ref{tab:deltaphi_q05}. The vertical dashed
    lines mark the time of formation of a common apparent horizon. The inset shows the
    early evolution of $|\varphi_c|$ for the heavy star B
    up to the point of
    divergence; plain superposition results in significant
    oscillations of $|\varphi_c|$. For $d/M=14.8$ and $29.7$,
    these oscillations trigger a premature
    collapse of the heavy star into a BH (signalled by the
    rapid drop of $|\varphi_c|$ to zero) well before a common
    apparent horizon forms. 
    }
\end{figure}
In this analysis of unequal-mass collisions, we encounter one minor difficulty, the {\it a-priori} unknown dephasing at merger.
To ensure that our comparison of binaries starting from
different initial separations is not adversely affected
by possible variations in the dephasing at merger,
we choose for each separation the initial phase-parameter that maximizes the radiated GW energy (cf.~Fig.~\ref{fig:energy_fit} below). For the mass ratio
$q=0.5$ and our two superposition types the specific values of the dephasing parameter can be found in Table \ref{tab:deltaphi_q05}. 
\begin{table}[hb!] \label{tab:deltaphi_q05}
\centering
\footnotesize
\begin{tabular}{C{3cm}|C{1cm}|C{1cm}|C{1cm}|C{1cm}}
  \hline
  $d/M$ & 14.8 & 29.7 & 44.5 & 59.3 \\
  \hline
  $\delta \phi$ (plain) & $330^{\circ}$ & $60^{\circ}$ & $60^{\circ}$ & $30^{\circ}$
  \\
  $\delta \phi$ (improved) & $240^{\circ}$ & $210^{\circ}$ & $120^{\circ}$ & $0^{\circ}$\\
  \hline
\end{tabular}
\caption{Choices of dephasing parameters $\delta \phi$ maximising the GW energy for $\texttt{q05-dX-pX}$ binary configuration.}
\end{table}

The resulting time evolutions $|\varphi_c(t)|$ are shown
for all four initial separations and both superposition
methods in Fig.~\ref{fig:phi_q05} and exhibit the same
features as mentioned above: plain superposition (left panel) results in
significant unphysical oscillations of $|\varphi_c|$
which for $d/M = 14.8$ and $d/M = 29.7$ also cause a collapse
of the heavier star B into a BH well before a common horizon
forms as marked by the vertical dashed lines. For the larger
separations $d/M=44.5$ and $d/M=59.3$, the premature BH
formation is avoided, but plain superposition still results
in significant pulsations of the BSs. In contrast, the
scalar-field amplitude of both BSs remains very close
to its equilibrium value for our improved superposition
method throughout the infall as demonstrated
in the right panel of Fig.~\ref{fig:phi_q05}. As expected, significant dynamics in the scalar field,
including the eventual collapse to a single BH, are only
encountered around merger: the local maximum in the scalar field
coincides with common horizon formation for small and large
initial separations alike.

We have repeated this analysis for different mass ratios
and different choices of the dephasing parameter.
As it turns out, the dephasing parameter has no significant
effect on the results shown in Fig.~\ref{fig:phi_q05} and
our above concern about choosing $\delta \phi$ appropriately
has been unnecessary. The mass ratio, however, does affect
the results to some extent. The spurious effects of plain
superposition are even more pronounced for $q=0.75$ (where
premature BH formation occurs for all but the largest $d/M$)
and less pronounced for $q=0.38$ (where only the smallest
initial separation results in premature BH formation).
This $q$ dependence is fully consistent with the above
observation (i) in Sec.~\ref{sec:results_unequal} that
plain superposition becomes viable for $q\rightarrow0 $.

\subsection{Gravitational waveforms}
\begin{figure}%
    \centering
    \includegraphics[width=\textwidth]{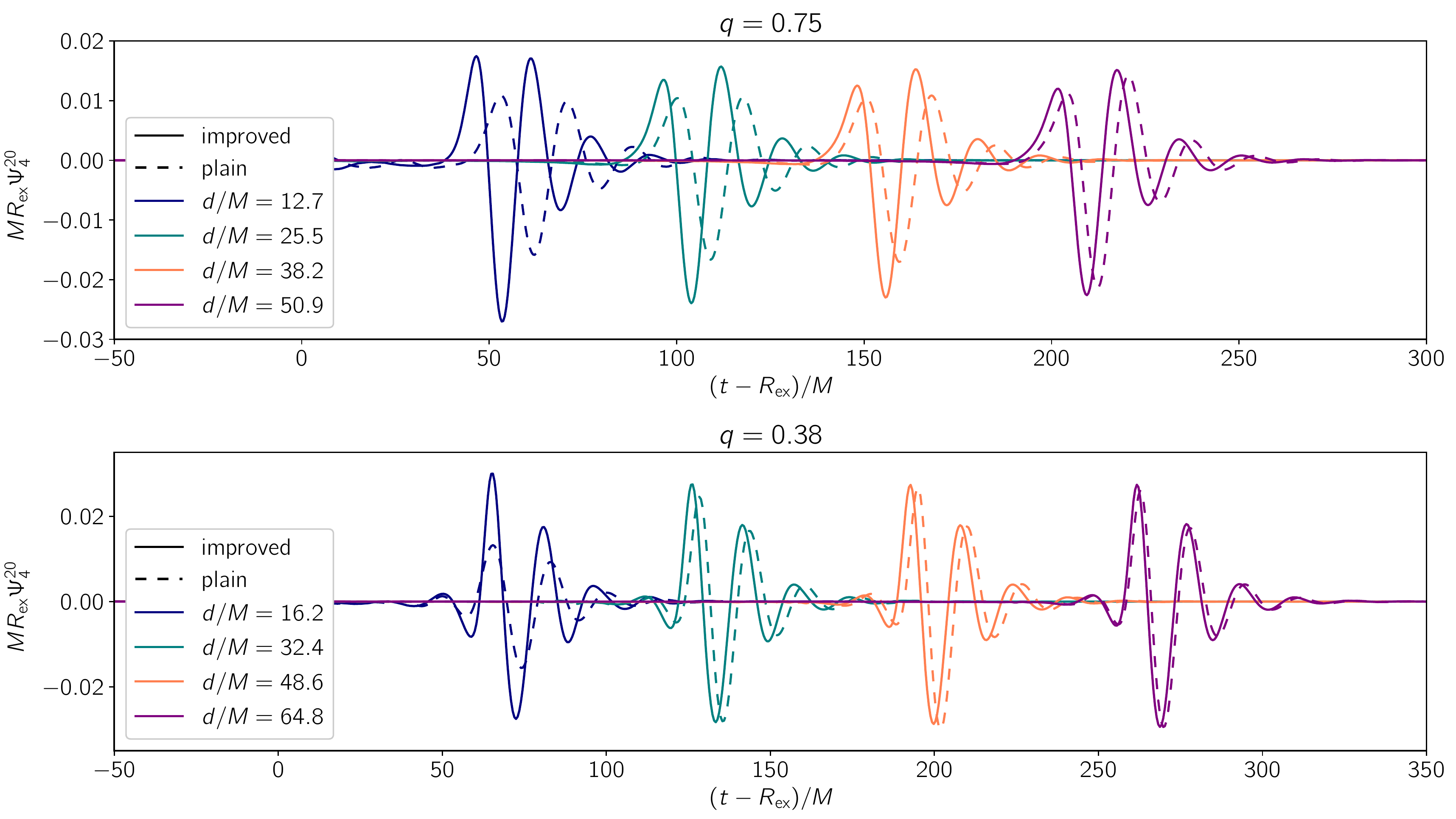}
    \label{fig:psi4_q075_q038}%
    \caption{$20$-modes of the Newman-Penrose scalar $\Psi_4$ of binary configurations with $q=0.75$ (upper) and $q=0.38$ (lower) with varying initial separation, $d$. For each mass ratio, initial separation and superposition method, we choose the phase off-set parameter that maximises the GW energy; cf.~Table \ref{tab:deltaphi_q075_q038}. The straight line shows the waveform obtained for our improved method, whilst the dashed line shows that for plain superposition. As we increase the separation, both superpositions give comparable results. This is most notable in the case of smaller mass ratio $q=0.38$, where good agreement is already reached at smaller distances $d/M\gtrsim 32.4$. 
    }
\end{figure}

As we have seen in Sec.~\ref{sec:equal_collisions} and,
in particular, in Fig.~\ref{fig:energy_fit_equal}, the
initial binary separation $d$ can have a significant
effect on the magnitude of the GW signal due to the
corresponding differences in the collision velocity around
merger. This variation arises additionally to the impact
due to the choice of the initial dephasing parameter
$\delta \phi$. We can still check the consistency of our
evolutions, however, in the limit of large separation $d$
and simultaneously selecting $\delta \phi$ such that
it maximizes the GW energy.
Specifically, we show the resulting GW signals
in Fig.~\ref{fig:psi4_q075_q038} for the $\texttt{q075-dX-pX}$ binaries with initial separation from $d/M=12.7$ to $d/M=50.9$ and the $\texttt{q038-dX-pX}$ binaries
with $d/M=16.2$ to $d/M=64.8$. The dephasing parameters $\delta \phi$ maximising the GW energy for these configurations are shown in Table \ref{tab:deltaphi_q075_q038}.
\begin{table}[hbt!] \label{tab:deltaphi_q075_q038}
\centering
\footnotesize
\begin{tabular}{C{3cm}|C{1cm}|C{1cm}|C{1cm}|C{1cm}|C{1cm}|C{1cm}|C{1cm}|C{1cm}}
  \hline
  & \multicolumn{4}{c|}{$q=0.75$} & \multicolumn{4}{c}{$q=0.38$} \\
  \hline
  $d/M$ & 12.7 & 25.5 & 38.2 & 50.9 & 16.2 & 32.4 & 48.6 & 64.8 \\
  \hline
  $\delta \phi$ (plain) & $300^{\circ}$ & $330^{\circ}$ & $330^{\circ}$ & $270^{\circ}$ & $330^{\circ}$ & $150^{\circ}$ & $180^{\circ}$ & $180^{\circ}$
  \\
  $\delta \phi$ (improved) & $180^{\circ}$ & $240^{\circ}$ & $270^{\circ}$ & $270^{\circ}$ & $180^{\circ}$ & $330^{\circ}$ & $300^{\circ}$ & $210^{\circ}$\\
  \hline
\end{tabular}
\caption{Choices of dephasing parameters $\delta \phi$ maximising the GW energy for $\texttt{q075-dX-pX}$ and $\texttt{q038-dX-pX}$ binary configurations.}
\end{table}

The Figure clearly demonstrates that for mass ratio $q=0.75$ $(q=0.38)$, our improved superposition method results in comparable GW amplitudes for initial separations $d/M \gtrsim 25.5$ $(d/M\gtrsim 32.4)$. This convergence in the GW signal (maximized over $\delta \phi$) is as expected for large initial separations. In contrast, plain superposition only achieves this convergence for much larger initial separations, namely $d/M \gtrsim 50.9$ $(d/M \gtrsim 32.4$). For small initial separations, plain superposition systematically results in weaker GW amplitude when compared to our improved superposition method. In fact, this is a distinct feature of premature BH formation that occurs
precisely for these plainly superposed configurations; cf.~Sec.~\ref{sec:Actr_profiles}. 
As expected, the agreement in the GW amplitude between the
two superposition methods improves as we increase the separation and/or decrease the mass ratio,
i.e.~the very limits described in items (i) and (ii) of Sec.~\ref{sec:results_unequal}.

\begin{figure}%
    \centering
    \includegraphics[width=15cm]{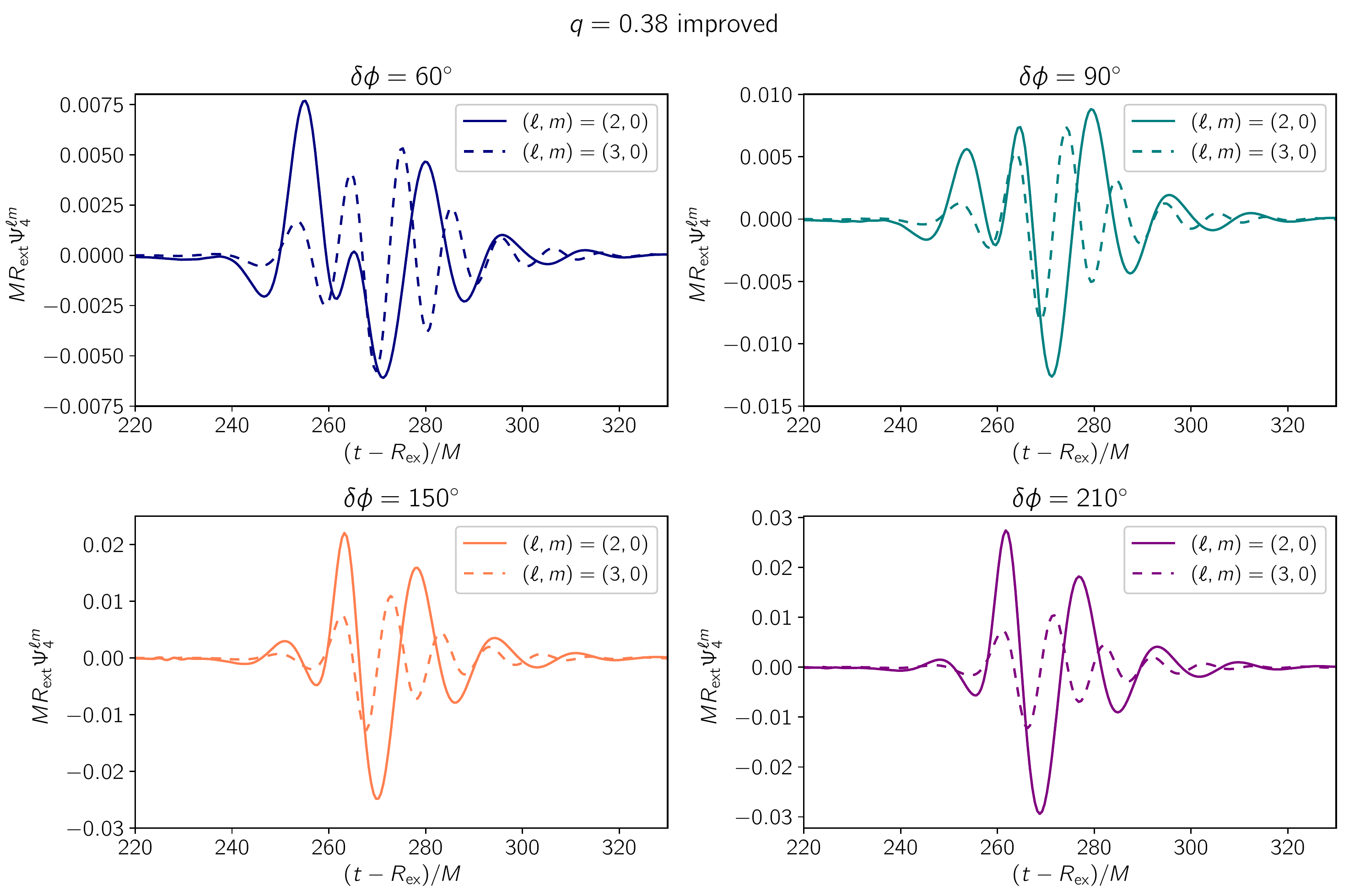}
    \label{fig:psi4_q038_deltaphi}%
    \caption{$20$-modes and $30$-modes of the Newman-Penrose scalar $\Psi_4$ for the improved binary configuration $q=0.38$ with initial separation $d/M=64.8$ and different phase off-set parameters $\delta \phi$. The prominent feature of certain off-set parameters (e.g. $\delta \phi = 60^{\circ}, 90^{\circ}$) is that they result in fainter GW signals with more complex structure and $\ell=3$ modes become comparable in amplitude to the $\ell=2$ ones. This is unlike the head-on BH case of the same mass ratio, where $\ell = 3$ mode is roughly 5 times smaller than the $\ell = 2$ mode \cite{Sperhake:2011ik}.
    }
\end{figure}

Quite remarkably, we obtain the largest GW amplitudes for
the smallest mass ratio $q=0.38$. This mass ratio also
exhibits the most interesting waveforms.
For a wide range of the dephasing angle $\delta \phi$, we find the GW radiation to be quadrupole dominated with a merger pulse
reminiscent of BH head-on collisions; see e.g.~Fig.8 of Ref.
\cite{Sperhake:2006cy}. However, certain phase off-set parameters result in fainter and aberrant GW signals with signatures of tidal deformation of the binary constituents. We attribute this more complex shape of the fainter signals to the lighter BS in the binary becoming more prone to tidal effects from its more compact and heavier companion. For a given self-interation constant $\sigma_0$ and with decreasing mass, it has been shown that tidal deformability of the star increases \cite{Sennett:2017etc}, and as a result the gravitational waveform can significantly depart from the BH-like form \cite{Johnson-Mcdaniel:2018cdu}.
We illustrate this phenomenon in more detail in Figure \ref{fig:psi4_q038_deltaphi}, which shows the $q=0.38$ waveforms for phase off-set parameters $\delta \phi = 60^{\circ}, 90^{\circ}, 150^{\circ}, 210^{\circ}$. The former two give the smallest GW amplitude and exhibit significant deformation, whilst the latter two show a clear, `black hole' like signal. In the case of fainter GW signals, the importance of higher modes becomes more prominent: as indicated in the same Figure, $\ell=3$ modes become almost comparable in amplitude to the $\ell=2$ modes for binaries with $\delta \phi = 60^{\circ}, 90^{\circ}$.

\subsection{Energy radiated by unequal-mass binaries}
Similar to the equal-mass binaries, in the unequal-mass case we also observe some discrepancy in the GW energy with varying initial separation and dephasing parameter. We recall that these effects (see Fig.~\ref{fig:energy_fit_equal}) have been attributed to the differences in the collision velocity and the degree of the repellent nature of the scalar field. However, in the unequal-mass case the dependency of the energy on the dephasing parameter and separation becomes even more complex. This is due to the fact that the phase difference at merger for unequal-mass binaries is no longer the initial dephasing $\delta \phi$ we apply to one of the BSs and as we vary the initial separation $d$ we further change the phase difference at merger. 

\begin{figure}%
    \centering
    \includegraphics[width=17cm]{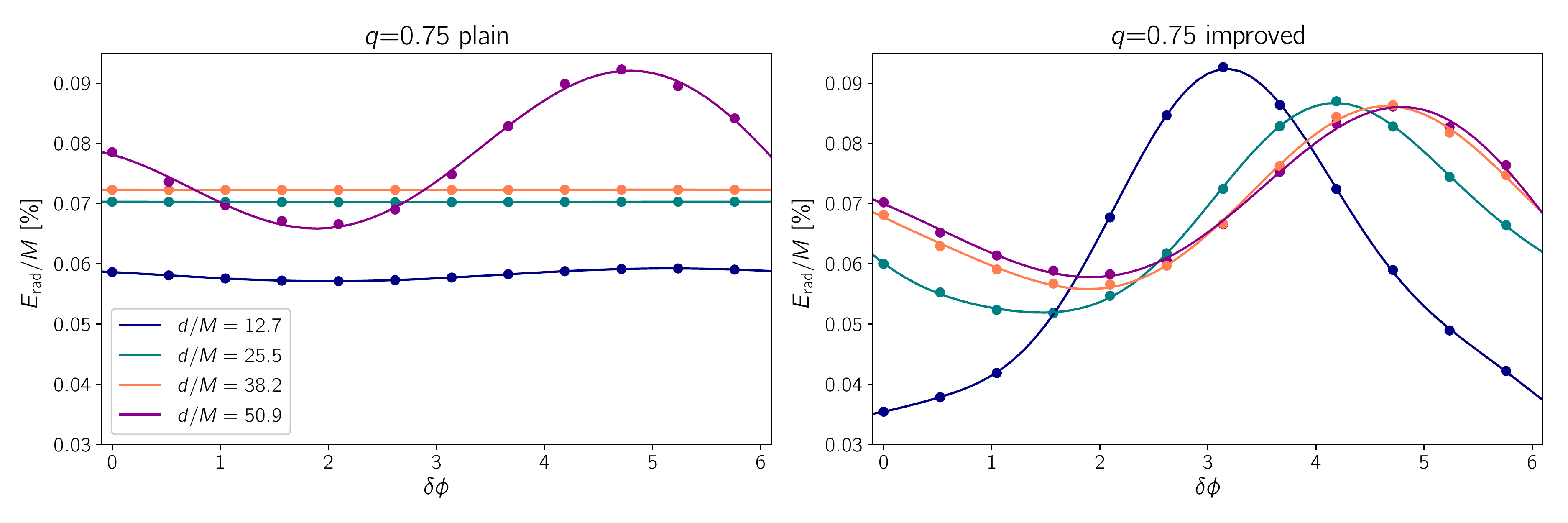}
    \includegraphics[width=17cm]{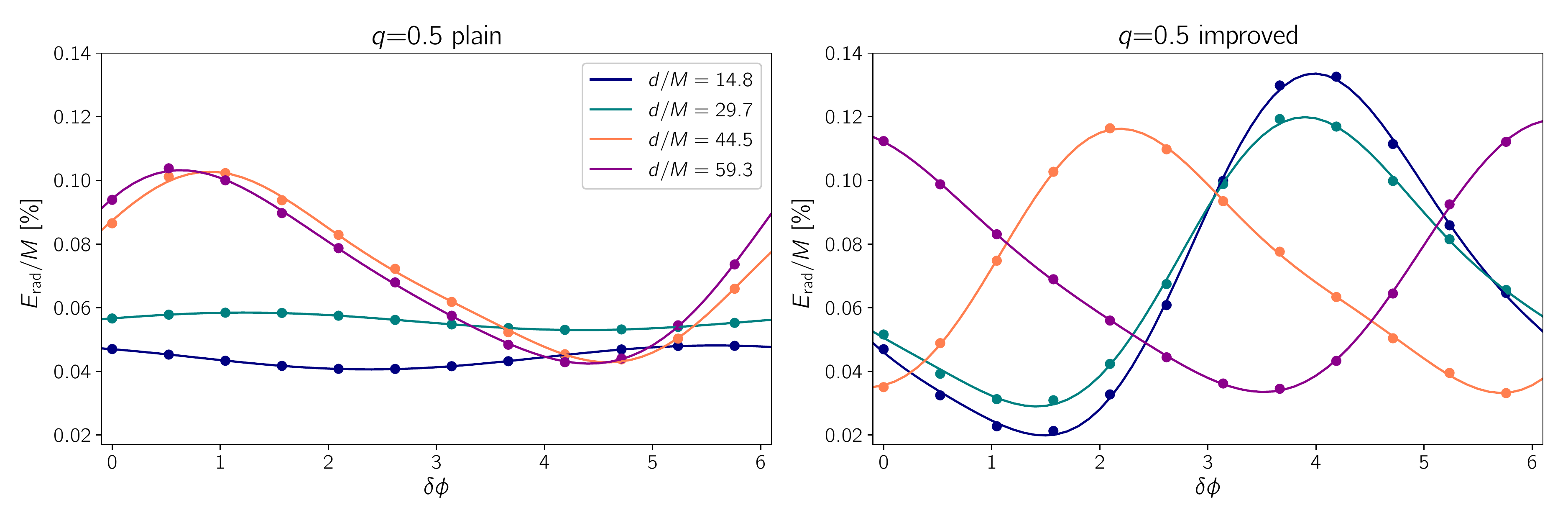}
    \includegraphics[width=17cm]{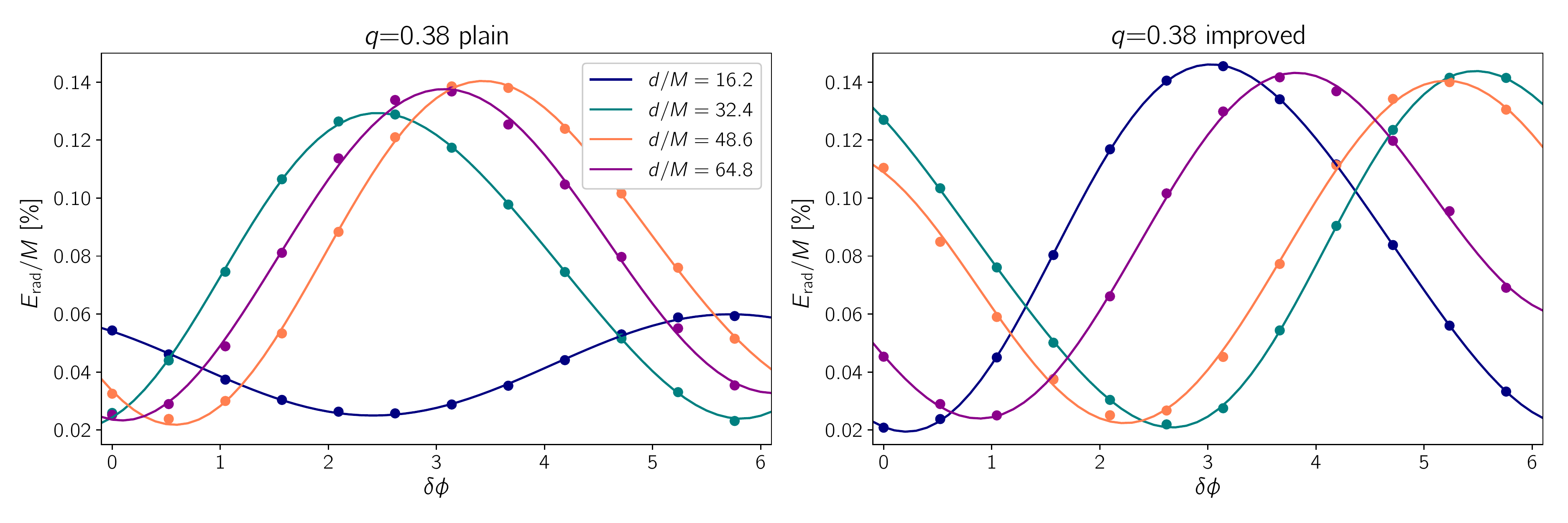}
    \caption{Radiated energy as a function of the phase off-set parameter $\delta \phi$ for binary collisions of $q=0.75,0.5,0.38$. The left panel shows the results for our improved superposition \eqref{eq:lambdanew} and the right panel those for plain superposition.
    }
    \label{fig:energy_fit}
\end{figure}

In Figure \ref{fig:energy_fit}, we illustrate this dependence
of the radiated energy on the dephasing parameter, $E_{\rm rad} (\delta \phi)$, for varying initial separations and all the mass ratios considered here. As shown in the left column, plain superposition results in flat energy profiles for smaller separations. We see here once again
a manifestation of premature BH formation which largely eliminates the effect of the scalar field's dephasing on the merger dynamics. 
This is unlike the energy profiles of our improved superposition displayed in the right column of Fig.~\ref{fig:energy_fit}. Here the energy dependence on the dephasing parameter takes on an approximately sinusoidal shape for all separations $d$. The distinct
horizontal shift between these profiles can be
attributed to the infall-time dependent
contribution to the dephasing $\delta \phi$.
Except for the smallest separation and modulo the horizontal shift, the energy profiles exhibit comparable maxima and minima as we change the initial separation.
However, similar to the equal-mass case, the results for
the smallest separation differ significantly,
presumably due to differing collision velocity as illustrated in Fig.~\ref{fig:transition}. 

Since the dependence of the radiated energy on $\delta \phi$ is time-dependent in the unequal-mass case, $E_{\rm rad}(\delta \phi)$ is no longer described by a single sinusoidal fit \eqref{eq:sinfit_1}. In fact, we find that a two-mode sinusoidal fit
\begin{equation}
    E_{\text{fit}} = \sum_{i=1}^2 A_i \text{sin}(f_i \delta \phi + p_i) + s
    \label{eq:sinfit_2}
\end{equation}
well approximates the data. This two-mode fit applies
to all configurations using the improved initial data
construction and the data of plain superposition at larger initial separations. Only in the case of small initial separations, where plain superposition results in premature BH formation, the energy is well fitted with the one-mode fit (\ref{eq:sinfit_1}); here the BH formation
eliminates the overall effect of the dephasing as well as any complications arising from its time dependence
during the merger stage. In summary, our results demonstrate that plain superposition not only results in quantitative changes in the emitted GW signals,
but also leads to a significant over-simplification of the merger dynamics.

\section{Conclusion} \label{sec:conclusion}

In this work, we have extended previous studies of boson-star (BS) binaries in two principle directions: (i) we have generalised the initial data construction for equal-mass BS binaries of Ref.~\cite{Helfer:2021brt} to the unequal-mass case, (ii) we have systematically explored the effect of the scalar field's dephasing $\delta \phi$ on the merger dynamics and GW emission.

For all mass ratios and dephasing parameters, we have shown how plainly superposed initial data can result in spurious physical effects, such as increased constraint violations, oscillations of the scalar fields' central amplitudes and premature BH formation. The key drawback of plain superposition that leads to these spurious effects is its failure to recover equilibrium values of the volume element at the stars' centres. By appropriately
correcting the conformal factor of the spatial metric,
our method
exactly recovers the equilibrium volume element
at the centres of the stars and thus circumvents the spurious features of plainly superposed data for all mass ratios and $\delta \phi$. Notably, our improved method explicitly incorporates the equal-mass fix and plain superposition as limiting cases.

Similar to previous studies (e.g.~\cite{Bezares:2017mzk,Palenzuela:2017kcg, Widdicombe:2019woy}), we find that the
choice of the dephasing parameter significantly affects
the merger dynamics of BS binaries, most notably through
the repellent character of merging scalar-field solitons with large phase differences. Crucially, for unequal-mass binaries, this phase difference is not equal to the initial dephasing parameter $\delta \phi$ but also
acquires a time-dependent contribution due to the individual BSs' different oscillation frequencies.

To leading order, we find that the radiated energy and, hence, GW amplitude depends approximately sinusoidally on
the dephasing parameter.

More specifically, we find equal-mass collisions to
result in the strongest GW signals (and maximal energy)
for $\delta\phi=0$, i.e.~configurations with equal phase
at merger. As the dephasing $\delta \phi$ is increased
towards $\pi$, 
the energy and GW amplitude decrease monotonically.
For unequal-mass binaries, we observe the same behaviour, accompanied, however, by a constant offset in $\delta \phi$ due to the differing oscillation frequencies and
the subsequent additional phase offset at merger;
cf.~Figs.~\ref{fig:energy_fit_equal} and \ref{fig:energy_fit}.

In general, the BSs' phase offset introduces considerable complexity to the merger dynamics and GW emission. In the equal-mass case, this manifests itself most prominently in a significant variation of the GW energy from off-phase BS binaries as we change the initial separation; cf.~Fig.~\ref{fig:energy_fit_equal}.
We attribute this variation to differences in the binaries' binding energies and the corresponding differences in the collision velocities as shown in Fig.~\ref{fig:transition}.
We observe the same phenomenon for unequal-mass binaries
but with maximal and minimal radiation occurring for shifted values of $\delta \phi$ which we ascribe to
the time dependent nature of the dephasing.
The extent of the energy discrepancy depends on the dephasing parameter. This may be connected to complex interactions between the collision velocity and the scalar fields' repulsion, which we leave for future study. 

Additionally our unequal-mass collisions exhibit several distinct features which we summarize as follows.
\begin{list}{\rm{{\bf(\arabic{count})}}}{\usecounter{count}
             \labelwidth0.5cm \leftmargin0.7cm \labelsep0.2cm \rightmargin0cm
             \parsep0.5ex plus0.2ex minus0.1ex \itemsep0ex plus0.2ex}
\item As shown in Fig.~\ref{fig:energy_fit}, smaller mass ratios produce larger GW energy than the equal-mass case. This is in contrast to the BH case, where $E_{\rm rad} \sim \eta^2 \defeq q^2 / (1+q)^4$ \cite{Berti:2007fi}, where $\eta$ is a monotonically decreasing function of $q$. 
\item As shown in Fig.~\ref{fig:psi4_q038_deltaphi}, certain dephasing parameters result in weaker GW signals with signatures of tidal deformation. For these configurations, higher modes, such as $\ell=3$, exhibit almost comparable
magnitude as their quadrupolar counterparts, especially in the case of binary with $q=0.38$. This is unlike the black hole case, where $\ell=3$ mode is 5 times smaller than the $\ell=2$ mode \cite{Sperhake:2011ik}.
\item The numerical data of the radiated energy $E_{\rm rad}(\delta \phi)$ in Fig.~\ref{fig:energy_fit} displays some deviation from the pure sinusoidal fit of \eqref{eq:sinfit_1} and is better described by the two-mode sinusoidal fit \eqref{eq:sinfit_2}. We believe this is a feature of the time-dependent phase difference \textit{during} the merger.
\end{list}

Clearly the enhanced parameter space leads to a rich structure in the merger dynamics of unequal-mass BS binaries which necessitates further systematic exploration, especially in the case of inspiralling
binaries. Furthermore, the proposed initial data superposition in this work is still an approximation to the ultimate goal of fully solving the constraint equations.
We leave these explorations for future efforts.

\ack
RC and TE are supported by the Centre for Doctoral Training
(CDT) at the University of Cambridge funded through STFC. TH is supported by NSF Grants No. AST-2006538, PHY-2207502, PHY-090003 and PHY20043, and NASA Grants No. 19-ATP19-0051, 20-LPS20- 0011 and 21-ATP21-0010.
This work has been supported by
STFC Research Grant No. ST/V005669/1
``Probing Fundamental Physics with Gravitational-Wave Observations''.
We acknowledge support by the DiRAC project
ACTP284 from the Cambridge Service for Data Driven Discovery (CSD3)
system at the University of Cambridge
and Cosma7 and 8 of Durham University through STFC capital Grants
No.~ST/P002307/1 and No.~ST/R002452/1, and STFC operations Grant
No.~ST/R00689X/1. We also acknowledge support by the DiRAC project grant DiRAC Project ACTP238 for use of Cosma7 and DiAL3. This research project was conducted using computational resources at the Maryland Advanced Research Computing Center (MARCC).
The authors acknowledge the Texas Advanced Computing Center (TACC) at The
University of Texas at Austin
and the San Diego Supercomputer Center for providing HPC resources that have contributed
to the research results reported within this paper through
NSF grant No.~PHY-090003. URLs: \url{http://www.tacc.utexas.edu}, \url{https://www.sdsc.edu/}.

\section*{References}
\bibliography{literature}
\bibliographystyle{unsrt}

\appendix

\clearpage

\section{Choice of conformal factor} \label{sec:conf_appendix}

In Eq.~(\ref{eq:conf_metric}), rewritten here as
\begin{equation}
  \gamma_{ij}=\lambda \tilde{\gamma}_{ij}
  ~~~~~\Leftrightarrow~~~~~
  \tilde{\gamma}_{ij}=\lambda^{-1}\gamma_{ij}
  ~~~~~\text{with}~~~~~
  \lambda=\gamma^{1/3}\,,
  \nonumber
\end{equation}%
we have expressed the conformal factor
in terms of the variable $\lambda=\gamma^{1/3}$. The exponent
$1/3$ is, of course, a free choice in this equation and we
will now explore the implications of choosing a different
exponent and why $1/3$ turns out to work particularly well in our practical applications.
Let us start with the conformal factor motivated by the Schwarzschild metric in (Cartesian) isotropic coordinates,
\begin{eqnarray} \label{eq:schwar_metric}
  \du s^2
  &=&
  -\left( \frac{2r-M}{2r+M} \right)^2 \du t^2
  + \left(1+\frac{M}{2r}\right)^4 \delta_{ij} \du x^i\,\du x^j
  \\[10pt]
  &=&
  -\left( \frac{2r-M}{2r+M} \right)^2 \du t^2
  + \psi^4 \delta_{ij} \du x^i\,\du x^j\,.
  \label{eq:Schwiso}
\end{eqnarray}
Note that the spatial metric is now written as
\begin{equation}
  \gamma_{ij} = \psi^4 \delta_{ij}\,,
\end{equation}
which is Eq.~(\ref{eq:conf_metric}) -- for the special case
of conformal flatness -- written in terms of the
alternative variable $\psi$ and an exponent $4$. We generalize
this freedom of writing the conformal factor by introducing
the variable
\begin{equation}
  \Lambda = \psi^n = \gamma^{n/12}~~~~~\text{with}~~~~~
  \gamma\defeq \det \gamma_{ij}\,,
\end{equation}
where\footnote{In principle we could allow for any $n\in\mathbb{R}$ here, but for practical reasons have performed numerical tests only for integer $n$.} $n \in \mathbb{Z}$.
For $n=-4$, for example, we recover the customary BSSN/CCZ4 conformal function $\Lambda= \chi$, whilst for the choice of $n=4$ we recover the conformal factor $\Lambda=\lambda = \gamma^{1/3}$ of Eq.~\eqref{eq:conf_metric}. In the general
case, we conformally decompose the spatial metric according to
\begin{equation}
  \gamma_{ij} = \gamma^{1/3} \tilde{\gamma}_{ij}
  = \Lambda^{4/n}\tilde{\gamma}_{ij}~~~~~\text{with}~~~~~
  \det \tilde{\gamma}_{ij}=1\,.
\end{equation}
In our construction of BS binary initial data,
we start with the spatial metric $\gamma_{ij}$ obtained
from plain superposition and then conformally
rescale this metric in order to correct the volume
element at the individual BSs' centers. This correction
gives us a new spatial metric
\begin{equation} \label{eq:general_n_rescaled_metric}
    \gamma^{\rm{new}}_{ij} = \left(\frac{\Lambda_{\rm{new}}}{\Lambda} \right)^{4/n} \gamma_{ij} = \frac{\Lambda_{\rm{new}}^{4/n}}{\gamma^{1/3}} \gamma_{ij}
    =\Lambda_{\rm new}^{4/n}\tilde{\gamma}_{ij}.
\end{equation}
By construction, this correction will recover the correct
volume element at the centers of both BSs for {\it any} choice
of $n$. The metric corrections thus introduced
{\it in the neighbourhood} of the BS centers, however, will differ for different choices of $n$. And we see in Eq.~(\ref{eq:general_n_rescaled_metric}) that
the conformal
rescaling is a linear function of our conformal variable
only for $n=4$. If we Taylor expand the rescaling factor
$\Lambda_{\rm new}/\Lambda$ around either BS center (where, we recall, it gives us the exact correction), then any choice
$n\ne 4$ will result in complicated additional terms in the
Taylor expansion of the factor $(\Lambda_{\rm new}/\Lambda)^{4/n}$.
We cannot rigorously prove that this non-linearity
in Eq.~(\ref{eq:general_n_rescaled_metric}) inevitably leads
to a significant deterioration of our initial data construction,
but this is exactly what we observe in all our numerical
experiments; choices $n\ne 4$ systematically result in significantly larger constraint violations compared to those
of Fig.~\ref{fig:constraints_q05} and generally more so the
further $n$ deviates from $4$.

As a summary, we list in Table \ref{conformal_factors}
the different variables for the conformal factor discussed in our initial data construction.
\begin{table}
\centering
\caption{\label{conformal_factors} A list of the conformal factor functions used in our discussion of binary initial-data construction. Note that in the main text we fix $n=4$ in the newly defined conformal factor, so that $\gamma_{ij}=\lambda_{\rm{new}}\tilde{\gamma}_{ij}$.
}
\footnotesize
\begin{tabular}{@{}lll}
\br
Name & Variable & Relation to $\tilde{\gamma}_{ij}$\\
\mr
Standard notation for isotropic Schwarzschild & \quad $\psi$ & $\gamma_{ij} = \psi^4 \tilde{\gamma}_{ij}=\psi^4\delta_{ij}$\\
General conformal function & $\Lambda$ = $\psi^n$ & $\gamma_{ij} = \Lambda^{4/n} \tilde{\gamma}_{ij}$\\
BSSN/CCZ4 conformal factor & \quad $\chi$ & $\gamma_{ij} = \chi^{-1} \tilde{\gamma}_{ij}$ \\

Conformal variable for $n=4$ (our preferred choice) &
\quad $\lambda$ & $\gamma_{ij} = \lambda \tilde{\gamma}_{ij}$ \\
\br
\end{tabular}\\
\end{table}

\section{Improved superposition in the limits $R_{\rm A}, R_{\rm B} \to 0$ or $R_{\rm A}, R_{\rm B} \to \infty$} \label{sec:limit_cases}

In Section \ref{sec:parameter_space_qneq1} we have
demonstrated that in the equal mass case our improved superposition
\eqref{eq:gamma_new}
with $R_A, R_B \to 0$ results in the same gravitational waveform  as
obtained with plain superposition, whilst with $R_A, R_B \to \infty$ we recover the waveform from the equal mass fix of Refs.~\cite{Helfer:2018vtq,Helfer:2021brt}. Here we derive
analytically that our proposed method indeed reduces to plain
superposition and the equal-mass fix in the respective limits.

\subsection{The limit $R_{\rm A}, R_{\rm B} \to 0$}
In the limit $R_{\rm A},R_{\rm B}\rightarrow 0$, the weight functions
of Eq.~(\ref{eq:weight_function}) become
\begin{equation}
  w_{\rm A}(x^i) = \frac{1}{r_{\rm A}}\,,~~~~~~~~~~
  w_{\rm B}(x^i) = \frac{1}{r_{\rm B}}\,,
\end{equation}
where $r_{\rm A}=||x^i - x^i_{\rm A}||$ and $r_{\rm B}=||x^i-x^i_{\rm B}||$. This implies, in particular, that in Eq.~(\ref{eq:hAB}) the
terms $w_{\rm A}(x^i_{\rm A})\rightarrow \infty$ and
$w_{\rm B}(x^i_{\rm B})\rightarrow \infty$ diverge at the respective stars' centres, whereas
$w_{\rm A}(x^i_{\rm B})$ and $w_{\rm B}(x^i_{\rm A})$ remain finite,
so that
\begin{equation}
  h_{\rm A}
  \rightarrow
  \frac{w_{\rm B}(x^i_{\rm B}) \delta \lambda(x^i_{\rm A})}
       {w_{\rm A}(x^i_{\rm A}) w_{\rm B}(x^i_{\rm B})}
  =
  \frac{\delta \lambda(x^i_{\rm A})}{w_{\rm A}(x^i_{\rm A})}
  \,,~~~~~
  h_{\rm B}
  \rightarrow
  \frac{w_{\rm A}(x^i_{\rm A}) \delta \lambda(x^i_{\rm B})}
       {w_{\rm A}(x^i_{\rm A}) w_{\rm B}(x^i_{\rm B})}
  =
  \frac{\delta \lambda(x^i_{\rm B})}{w_{\rm B}(x^i_{\rm B})}\,.
\end{equation}
The correction $\delta \lambda(x^i)$ applied to the conformal
factor in Eq.~(\ref{eq:lambdanew}) then becomes
\begin{equation}
  \delta\lambda(x^i)
  =
  w_{\rm A}(x^i) h_{\rm A} + w_{\rm B}(x^i) h_{\rm B}
  = 
  \begin{cases}
  \delta \lambda(x^i_{\rm A})~~~~~& \text{for } x^i= x^i_{\rm A} \\
  \delta \lambda(x^i_{\rm B})~~~~~& \text{for } x^i= x^i_{\rm B} \\
  0 & \text{otherwise}
  \end{cases}\,.
\end{equation}
We thus recover $\delta \lambda = 0$, i.e.~plain superposition, everywhere except
at the isolated points $x^i_{\rm A}$ and $x^i_{\rm B}$. The
Dirac $\delta$ function like correction at $x^i_{\rm A}$ and $x^i_{\rm B}$ is a consequence of our condition (\ref{eq:conditionAB_2})
but is lost in numerical evolutions due to finite resolution,
so that for $R_{\rm A}, R_{\rm B}\rightarrow 0$ we expect to
obtain the same results as for plain superposition.

\subsection{The limit $R_{\rm A}, R_{\rm B} \to \infty$}
For $R_{\rm A}, R_{\rm B}\rightarrow \infty$, we can write
the weight functions (\ref{eq:weight_function}) as
\begin{equation}
  w_{\rm J}(x^i) = \frac{1}{\sqrt{R_{\rm J}^2+ r_{\rm J}^2}}
  = \frac{1}{R_{\rm J}} \left(1+\frac{r_{\rm J}^2}{R_{\rm J}^2} \right)^{-1/2}
  \approx
  \frac{1}{R_{\rm J}} \left(1-\frac{r_{\rm J}^2}{2R_{\rm J}^2} \right)
  \,,
\end{equation}
where ${\rm J} = {\rm A}, {\rm B}$ and we have Taylor expanded to first order in $\frac{r_{\rm J}^2}{R_{\rm J}^2}$. Bearing in mind that
$r_{\rm A}(x^i_{\rm A}) = r_{\rm B}(x^i_{\rm B})=0$ and
$r_{\rm A}(x^i_{\rm B}) = ||x^i_{\rm B}-x^i_{\rm A}||
=r_{\rm B}(x^i_{\rm A})$ is simply the separation $d$ of the two
BSs, we obtain, again to leading order,
\begin{align}
  w_{\rm A}(x^i_{\rm A}) w_{\rm B}(x^i_{\rm B})
  - w_{\rm A}(x^i_{\rm B}) w_{\rm B}(x^i_{\rm A})
  &\approx \frac{1}{R_{\rm A}R_{\rm B}}
  - \frac{1}{R_{\rm A}R_{\rm B}}
  \left( 1-\frac{d^2}{2R_{\rm A}^2} \right)
  \left( 1-\frac{d^2}{2R_{\rm B}^2} \right)
  \nonumber \\[10pt]
  &
  \approx
  \frac{1}{R_{\rm A}R_{\rm B}}
  \left( \frac{d^2}{2R_{\rm A}^2} + \frac{d^2}{2R_{\rm B}^2} \right)\,.
\end{align}
We likewise expand, to leading order in $\frac{r_{\rm J}^2}{R_{\rm J}^2}$,
the coefficients $h_{\rm A}$, $h_{\rm B}$ in the
form of their linear combination in Eq.~(\ref{eq:lambdanew}),
\begin{align}
  &w_{\rm A}(x^i) h_{\rm A} + w_{\rm B}(x^i) h_{\rm B}
  \nonumber \\[10pt]
  &~~
  \approx \frac{
  \left[
  -\frac{1}{R_{\rm B}}
  \left(
  1-\frac{d^2}{2R_{\rm B}^2}
  \right)
  \delta\lambda(x^i_{\rm B})
  + \frac{1}{R_{\rm B}}\delta\lambda(x^i_{\rm A})
  \right]
  \frac{1}{R_{\rm A}}
  +
  \left[
  \frac{1}{R_{\rm A}}\delta\lambda(x^i_{\rm B})
  -\frac{1}{R_{\rm A}}
  \left(
  1-\frac{d^2}{2R_{\rm A}^2}
  \right)
  \delta\lambda(x^i_{\rm A})
  \right]
  \frac{1}{R_{\rm B}}
  }
  {w_{\rm A}(x^i_{\rm A}) w_{\rm B}(x^i_{\rm B}) -w_{\rm A}(x^i_{\rm B})w_{\rm B}(x^i_{\rm A})}
  \nonumber \\[10pt]
  &~~
  =
  \frac{
  -\left(
  1-\frac{d^2}{2R_{\rm B}^2}
  \right)
  \delta\lambda(x^i_{\rm B})
  +\delta\lambda(x^i_{\rm A})
  +\delta\lambda(x^i_{\rm B})
  -\left(
  1-\frac{d^2}{2R_{\rm A}^2}
  \right)
  \delta \lambda(x^i_{\rm A})
  }
  {\frac{d^2}{2R_{\rm A}^2} + \frac{d^2}{2R_{\rm B}^2}}
  \nonumber \\[10pt]
  &~~
  = \frac{
  \frac{d^2}{2R_{\rm B}^2} \delta\lambda(x^i_{\rm B})
  +\frac{d^2}{2R_{\rm A}^2} \delta\lambda(x^i_{\rm A})
  }
  {\frac{d^2}{2R_{\rm A}^2} + \frac{d^2}{2R_{\rm B}^2}}\,.
\end{align}
In the equal-mass case, the corrections at the BSs' centers are the same,
$\delta\lambda(x^i_{\rm A})=\delta \lambda(x^i_{\rm B})$, so that
\begin{equation}
  w_{\rm A}(x^i) h_{\rm A} + w_{\rm B}(x^i) h_{\rm B}
  =
  \delta \lambda(x^i_{\rm A}) = \delta \lambda(x^i_{\rm B})\,,
\end{equation}
which, after multiplication with $\tilde{\gamma}_{ij}$,
is the equal mass fix (\ref{eq:equalmassfix}).

\section{Gravitational radiation and numerical uncertainties}\label{sec:appendix_convergence}
In this section, we calibrate the accuracy of our numerical
simulations by studying the convergence of two BS binary
configurations, one obtained with {\sc grchombo} and one
with the {\sc lean} code.
%
%
\subsection{Extraction of physical quantities}
\label{sec:physical_quantites}
In our convergence studies for both codes we use the radiated energy.
For this purpose, we extract the GW signal in the form
of the Newman-Penrose \cite{Newman:1961qr, Bishop:2016lgv} scalar $\Psi_4$ for outgoing
radiation as described in Appendix A of Ref.~\cite{Radia:2021smk}. We decompose $\Psi_4$
into spin-weight $s=-2$ spherical harmonics according
to
\begin{equation} 
   \label{eq:psi4}
   \Psi_{4, lm} (t, R_{\rm ex}) 
   = 
   \int_{S^2} \:
   \Psi_4(t, R_{\rm ex}, \theta, \phi) \:
   \overline{Y^{-2}_{lm}} (\theta, \phi) \text{d}\Omega,
\end{equation}
where $S^2$ is a 2-sphere of fixed coordinate radius 
$R_{\rm{ex}}$, $Y^{-2}_{lm}$ are spin-weighted spherical harmonics~\cite{Brugmann:2008zz} and $\rm{d} \Omega = \sin \theta \rm{d}\theta \rm{d}\phi$.
We also compute the radiated energy via
\begin{equation}
    E_{\rm{rad}} (t) 
   = 
   \lim_{r \to \infty} \frac{r^2}{16 \pi} 
   \int_{t_0}^{t} dt' \oint_{S^2} d\Omega \:
   \mathbf{e}_r \left|\int_{-\infty}^{t'} dt'' \Psi_4 \right|^2
   , \label{eq:E_rad_GW}
\end{equation}
where $\mathbf{e}_r$ is the unit radial vector of a sphere.
\begin{figure}[t]
  \centering
  \includegraphics[width=350pt, clip=True]{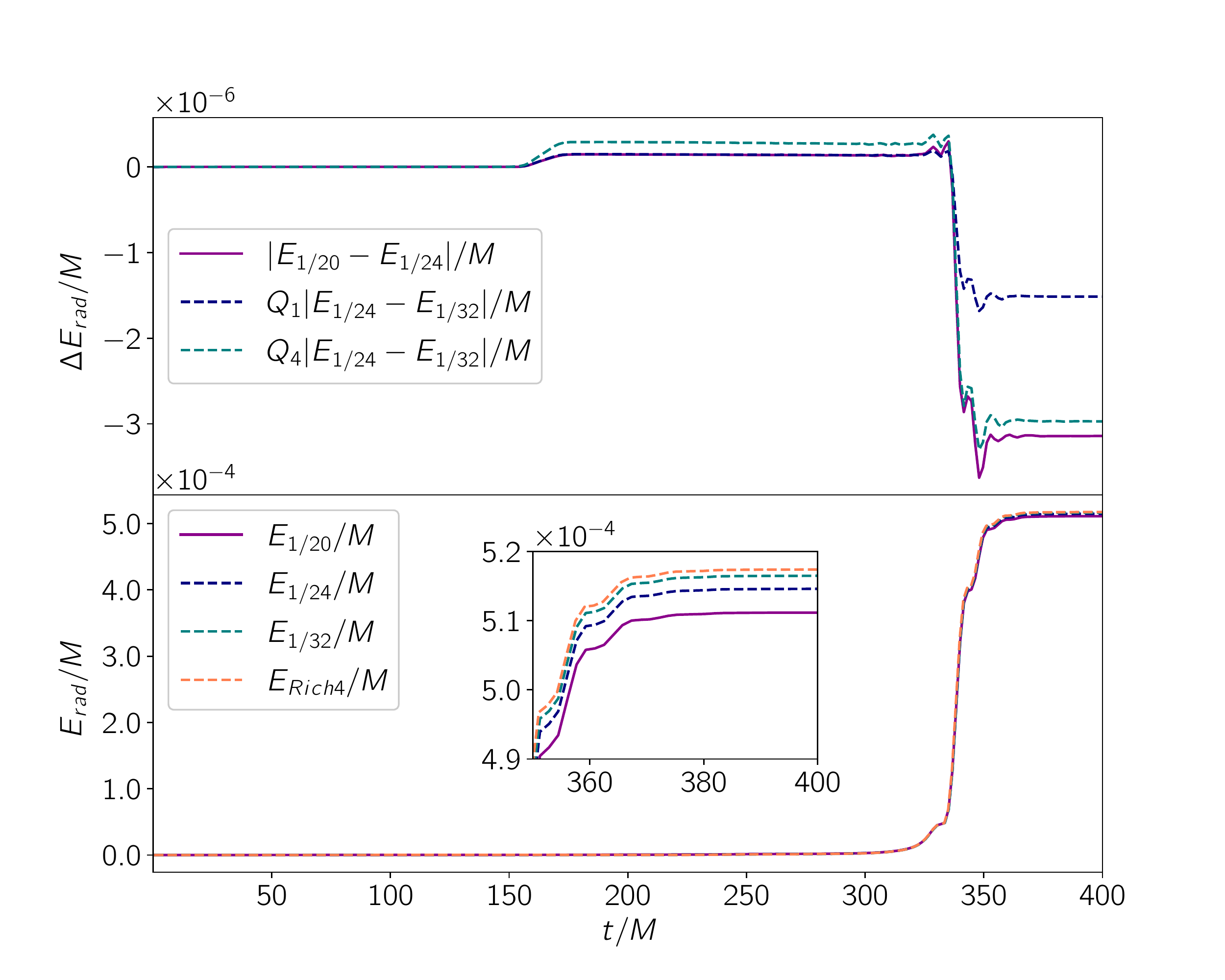}
  \caption{Convergence analysis of the GW energy computed with
  the {\sc Lean} code for the collision
  and merger of the BS binary {\tt q05\_d29\_p000} of Table \ref{configurations}. The upper panel shows the
  differences in the energy results obtained for different
  resolutions, measured here in terms of the grid spacing
  on the innermost refinement level. The high-resolution
  difference has been rescaled by $Q_1=0.8$ and $Q_4=1.57$
  corresponding to first and fourth-order convergence.
  While the small contribution due to noisy junk radiation
  converges only at first order, the total radiated energy
  exhibits convergence close to fourth order. The bottom
  panel displays the radiated energy as a function of time for
  the three resolutions as well as a fourth-order Richardson
  extrapolation. 
  }
  \label{fig:convEradq20}
\end{figure}
\subsection{{\sc lean} code}
The first case we study is the binary configuration
{\tt q05\_d29\_p000} of Table \ref{configurations}. In terms of the notation of
section \ref{sec:leancode}, the grid setup for these runs
is given by a domain size $L_1=1024$
with grid spacing (on the innermost refinement level) $\du x_7=1/20$, $1/24$ and $1/32$,
respectively, for a total mass $M=1.0787$. In figure \ref{fig:convEradq20}, we show
the resulting convergence analysis for the energy $E_{\rm rad}$
radiated in GWs. We see from the upper panel of the figure
that the total radiated energy converges at about fourth order.
The early part of the signal, which is dominated by the
high-frequency contributions from the spurious ``junk''
radiation converges at lower order, approximately first,
but does not significantly affect the total radiated energy.
Through comparison with the Richardson extrapolated values,
we estimate the discretization in the radiated energy to be
about $0.6\,\%$ at medium resolution $\du x_7=1/24$ which
is the resolution used in our {\sc lean} production runs.

The second main uncertainty in our results arises from the
extraction of the GW signal at finite radius. We determine
this error by extrapolating the GW signals computed
at seven equidistant extraction radii in the
range $R_{\rm ex}=120$ to $240$ using a first order fit
in $1/R_{\rm ex}$ as described in section 4.13
of \cite{Radia:2021hjs} and obtain a numerical uncertainty
of $1.4\,\%$ for the radiated energy. Combined with the
discretization error, this gives us a numerical error budget
of $2\,\%$.

We have analyzed in the same way the quadrupole of the
Newman-Penrose scalar $\Psi_4$ and observe the same order
of convergence but find a total error about twice as large,
$4\,\%$ in the GW amplitude.

\subsection{{\sc GRChombo} code}
\begin{figure}
  \centering
  \includegraphics[width=250pt]{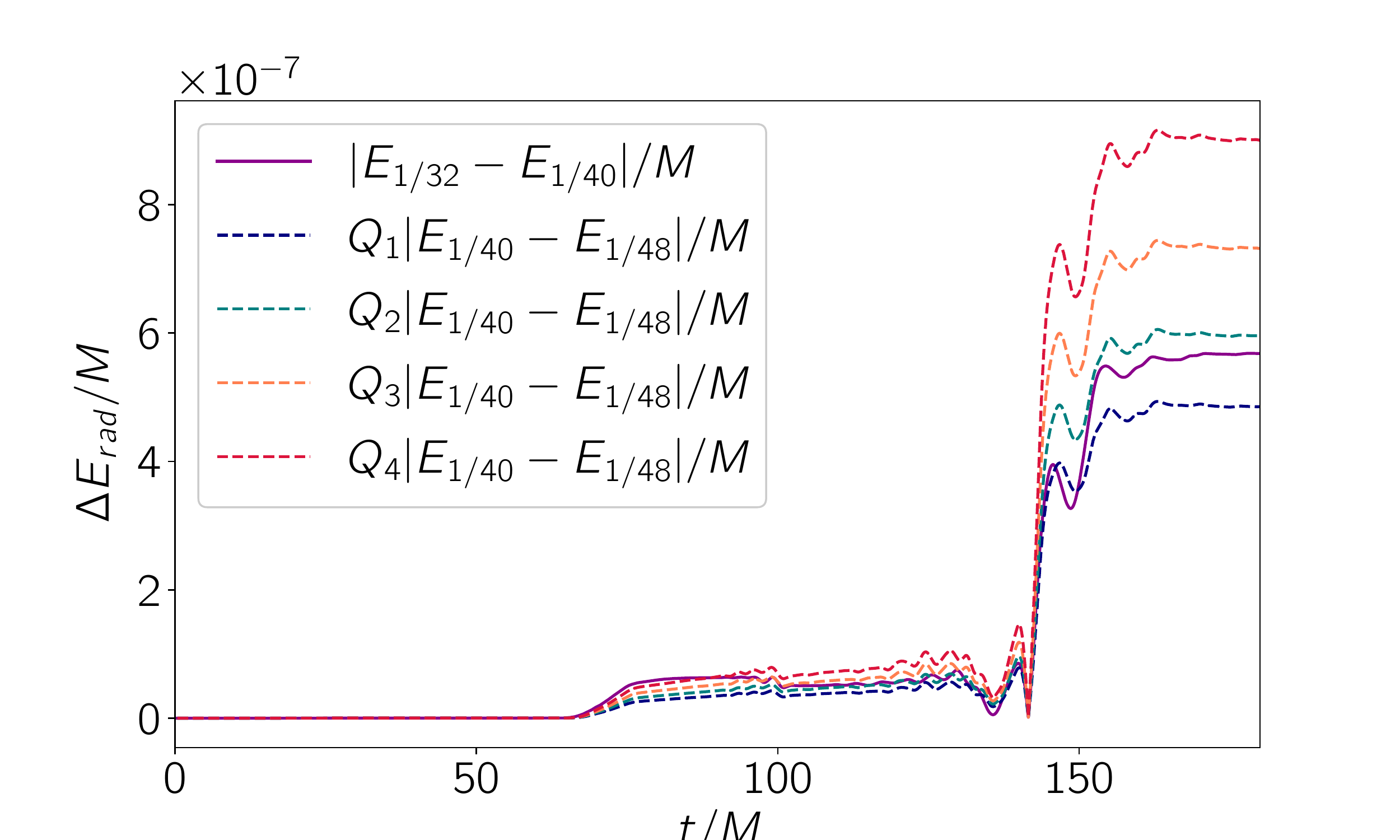}
  \caption{Convergence analysis of the GW energy computed with
  at $R_{\rm{ex}} = 120$ for the {\sc GRChombo} simulation of BS binary {\tt q075\_d12\_p000} of Table \ref{configurations}. The difference between high and medium resolutions has been rescaled by factors $Q_1$ to $Q_4$
  corresponding to convergence of first to fourth order.
  At early times, the order of convergence fluctuates between second and third, whilst at later times it settles to just below second order of convergence.
  }
  \label{fig:convEradq075}
\end{figure}

Here we study the binary configuration {\tt q075\_d12\_p000} of Table \ref{configurations}. We use the same grid set-up as described in the main section with $\du x = 1/32$, $\du x = 1/40$ and $\du x = 1/48$ on the finest refinement level for low, medium and high resolutions respectively; here the total mass is $M=1.2558$. The results shown in Fig.~\ref{fig:convEradq075} demonstrate overall second-order convergence. By comparison with the Richardson extrapolated results, we obtain a discretization error of around $1.1 \%$ for the low resolution, on which the results of the main text are based on. 

We estimate the error due to the finite extraction radius by extrapolating the GW signals computed at six equidistant radii in the range $R_{\rm{ex}} = 60 - 120$ and find an uncertainty of $2.6 \%$.
This thus gives us a total error budget for {\sc GRChombo} of $3.7 \%$.

\section{Constraint violations in the $(R_{\rm A}, R_{\rm B})$ parameter space} \label{sec:heat_maps_appendix}

In Figure \ref{fig:heat_map_q075} of Section \ref{sec:parameter_space_qneq1}, we have shown the behaviour of the $L2$ norm of Hamiltonian constraint over the parameter space $(R_{\rm A}, R_{\rm B})$. In this Appendix we attach results for the other unequal-mass binary configurations with $q=0.5$ and $q=0.38$.

\begin{figure}[ht] 
  \label{ fig7} 
  \begin{minipage}[b]{0.5\linewidth}
    \centering
    \includegraphics[width=\linewidth, valign=c]{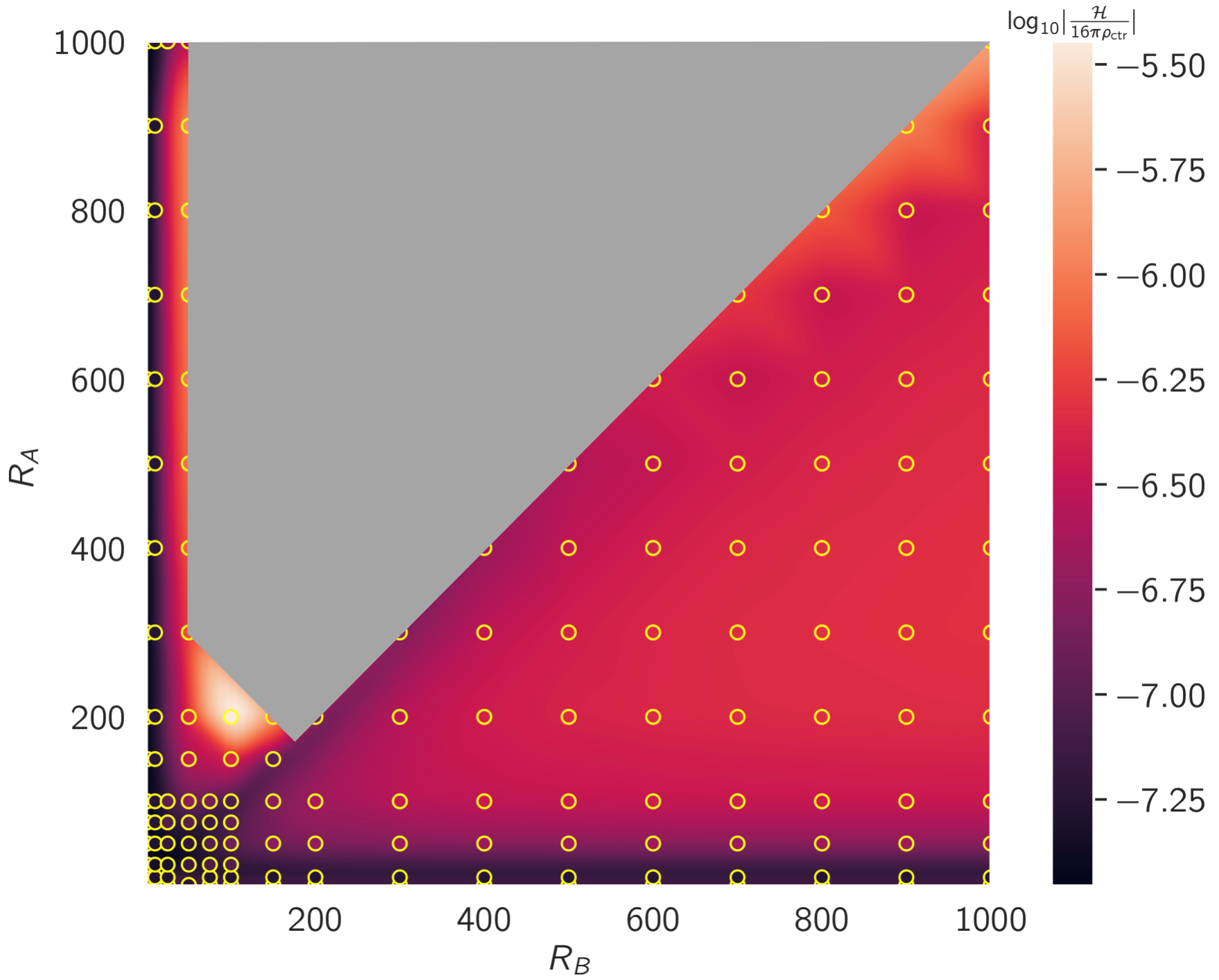} 
    \vspace{2ex}
  \end{minipage}
  \begin{minipage}[b]{0.5\linewidth}
    \centering
    \includegraphics[width=0.97\linewidth, valign=c]{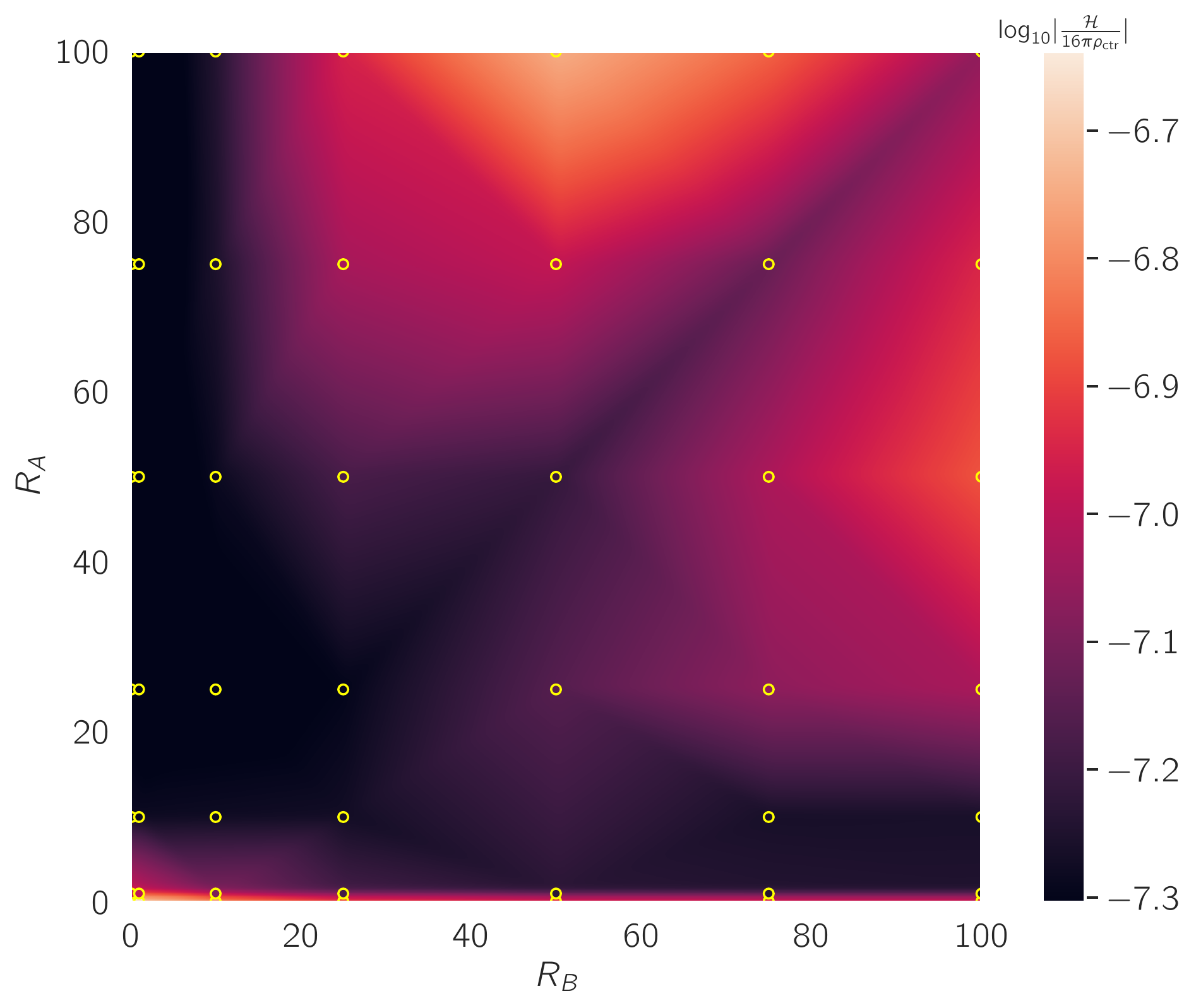} 
    \vspace{2ex}
  \end{minipage} 
  \begin{minipage}[b]{0.5\linewidth}
    \centering
    \includegraphics[width=0.99\linewidth, valign=c]{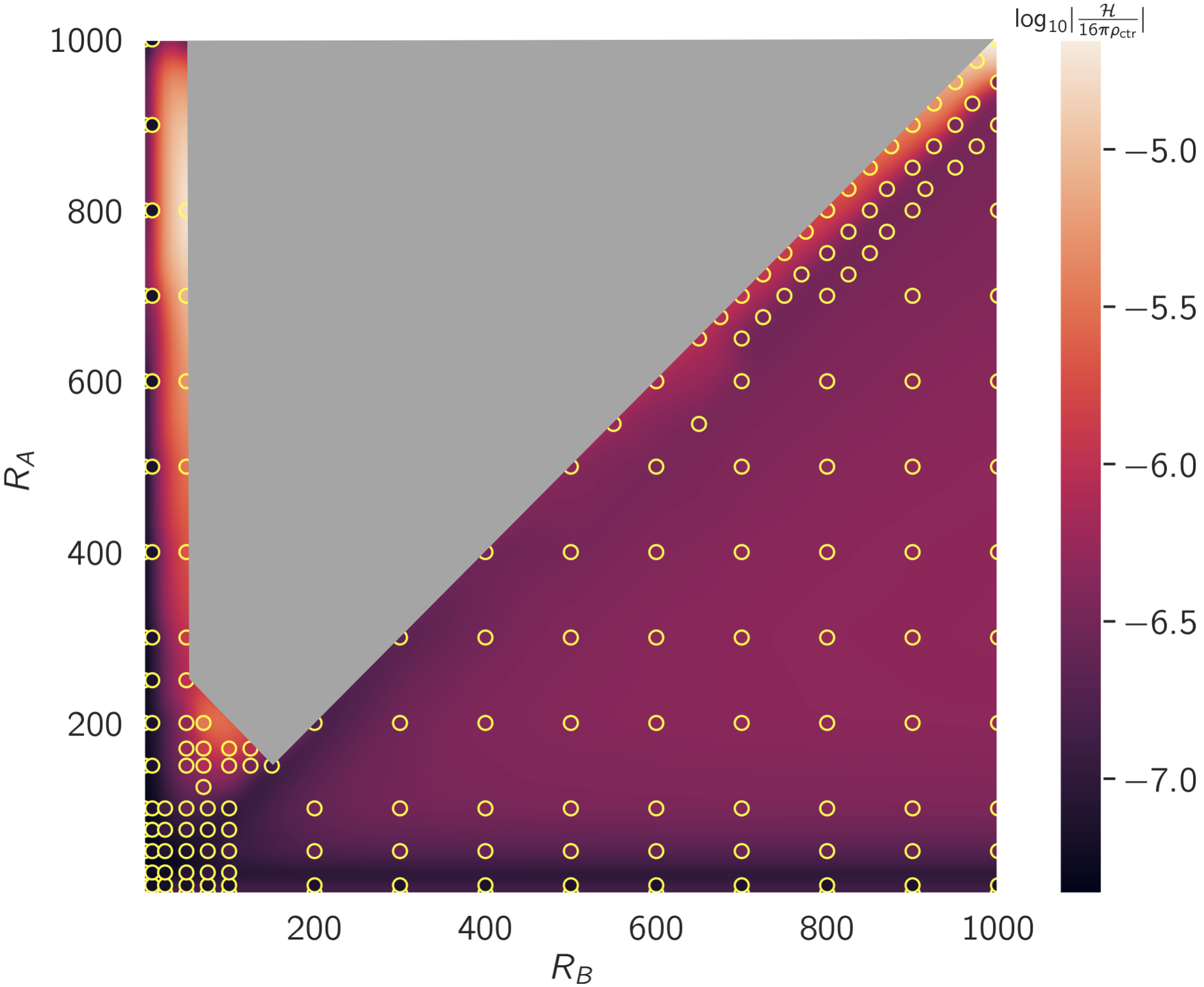} 
    \vspace{2ex}
  \end{minipage}
  \begin{minipage}[b]{0.5\linewidth}
    \centering
    \includegraphics[width=0.97\linewidth, valign=c]{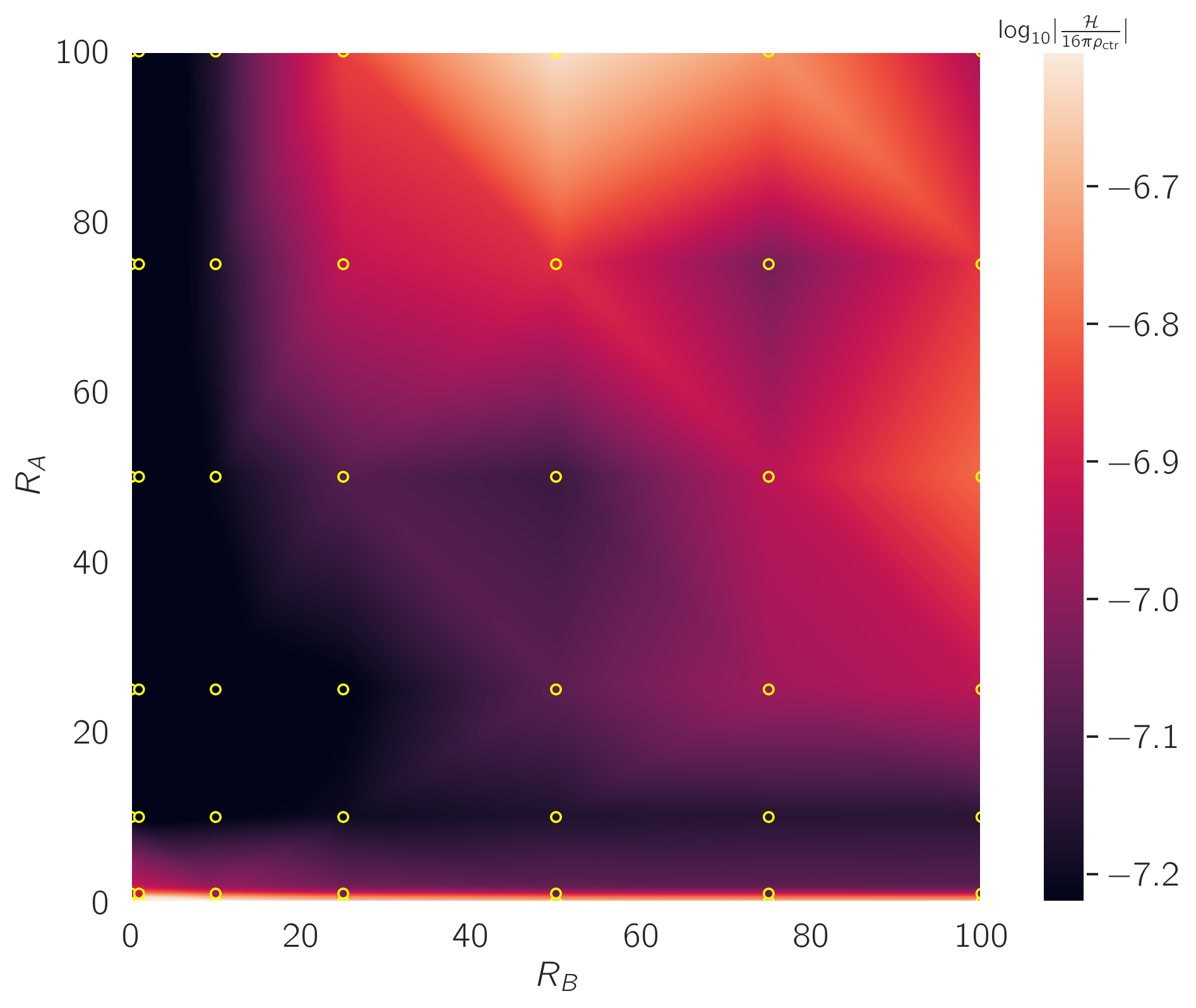} 
    \vspace{2ex}
  \end{minipage} 
  \caption{The logarithm of the $L2$-norm of the Hamiltonian constraint violations is shown for the binary configurations {\tt q05-d14-p000} (upper) and {\tt q038-d16-p000} (lower panels). For each case, the right panel shows a zoom-in on the region $(R_{\rm{A}}, R_{\rm{B}}) = [(0,100) \times (0,100)]$. The grey region indicates the parameter space of $(R_{\rm{A}}, R_{\rm{B}})$, where constraint violations are too large; note that this region increases in size for smaller mass ratios.} \label{fig:heatmaps_q05q038}
\end{figure}

\end{document}